\documentstyle[12pt]{article}
\newtheorem{theorem}{Theorem}
\newtheorem{proposition}{Proposition}
\newtheorem{definition}{Definition}
\newcommand{ \R} {\mbox{\rm I$\!$R}}
\begin{document}

\title{Nonlinear Connections and Nearly\\ Autoparallel Maps in General Relativity}
 \author{Heinz Dehnen\thanks{e--mail:\
 Heinz.Dehnen@uni-konstanz.de}\quad and  Sergiu I.\ Vacaru  \thanks{ e--mail:\
vacaru@lises.asm.md, on leave of absence from
 the Institute of Applied Physics,\newline Academy of Sciences,
  Chi\c sin\v au MD2028, Republic of Moldova } \quad
    \\[6pt]\\
\small  Fachbereich Physik, Universitat Konstanz, \\ \small
Postfach M 677, D--78457, Konstanz, Germany
   }
\date{August 15, 2000}
\maketitle

\begin{abstract}
We apply the method of moving anholonomic frames, with associated
nonlinear
 connections, in (pseudo) Riemannian spaces and examine the conditions when
various types of locally anisotropic (la) structures (Lagrange,
Finsler like and more general ones) could be modeled in general
relativity. New classes of solutions of the Einstein equations
with generic local anisotropy are constructed.  We formulate
 the theory of nearly autoparallel (na) maps and introduce the tensorial na--integration
as the inverse operation to both covariant derivation and
deformation of connections by na--maps. The problem of
redefinition of the Einstein gravity theory
 on na--backgrounds, provided with a set of na--map invariant conditions and
 local conservation laws,   is analyzed.  There are illustrated some examples of generation
of vacuum Einstein fields by Finsler like metrics and chains of
na--maps.
\end{abstract}



\section{Introduction}

It is possible that a corresponding spacetime anholonomic frame
structure
 induces local an\-isot\-ro\-pies which also 'slightly' modifies the isotropy of cosmic background
radiation. Usually, by developing different cosmological scenaria,
it is believed that  anisotropies and inhomogeneities
 in the Universe are caused only by  anisotropic matter (classical
and/or quantum) fluctuations and distributions.  To compare some
anisotropic effects of two different origins (induced
geometrically  by imposing some anholonomic conditions
 or by matter fields anisotropies) it is necessary to perform a rigorous
 definiton and analysis of physical values and fundamental field equations with
 respect to such anholonomic bases when there are modeled constrained dynamical systems with 'mixed'
  holonomic and  anholonomic variables.

Such models of strings, gravity and matter field locally
anisotropic (la) interactions (in brief, we use such terms as
la--gravity, la--strings, la--spacetime and so on) have been
proposed with the aim to illustrate that the so--called 'locally
anisotropic physics' follows alternatively in the low energy
energy limits of (super) string theory and that self--consistent
theories with both la--spacetime  and la--matter could be
constructed \cite{vjmp,vap,vnp,vjhep}. There are developed
different approaches to la--spacetimes and la--gravity which are
grounded on Finsler geometry and generalizations
\cite{fin,car35,rund,mat,as,asp,ma87,ma94,bej}. For instance,
there is a subclass of models with local anisotropy which try to
describe possible violations of local Lorentz invariance in the
framework of Finsler geometry \cite{bog1,bog2,goenner}. This non
infrequently introduces the misunderstanding that an 'unusual'
relativity is presented in all Finsler like theories and lied to
the misinterpretation that experimentally such Finsler spaces met
rather stringent constrains \cite {will}.

A surprising result is that Finsler like metrics and their
generalizations could be found as solutions of the Einstein
equations in general relativity and higher dimension gravity
models (see \cite{v2000} and Sections 2 and 8 in this work). The
point is to model various type of la--structures (Finsler type and
more general ones) by using anholonomic frames on (pseudo)
Riemannian spacetimes. This class of la--spacetimes are compatible
with the paradigm of the Einstein--Lorenz--Poincare relativity and
the theory of experiment on such curved spacetimes have to be
adapted to the viewpoint of observers stated with respect to
anholonomic frames of reference.

The problem of equivalence of spaces with generalized metrics and
connections was considered in a series of works by E. Cartan \cite
{car23,car30,car35a,car35,car39,car45} who developed an unified
approach to Riemannian, affine and projective connection, fiber
bundles, Finsler and another type of curved spaces on the language
of moving frames and differential forms. Paper \cite{car35}
contains also the idea on nonlinear connection (N--connection)
associated to an anholonomic frame. The global definition of
N--connection is due to W. Barthel \cite{barth} and this concept
was developed and applied by R. Miron and M. Anastasiei
\cite{ma87,ma94} in their geometry of generalized Lagrange and
Finsler spaces modeled on vector and tangent bundles. The further,
in this line, geometric extensions and applications in physics are
connected with the (super) frames, metrics and connections in
spinor spaces and superbundles provided with N--connection
structures and the geometry of locally anisotropic strings and
gravity \cite {vjmp,vjhep,vap,vnp}.

The first purpose of this paper is to demonstrate that anholonomic
frame structures with associated N--connections on (pseudo)
Riemannian spacetimes display a new 'locally anisotropic' picture
of the Einstein gravity. Here is to be noted that the elaboration
of models with la--interactions is considered to entail great
difficulties because of the problematical character of the
possibility and manner of definition of conservation laws on
la--spaces. It will be recalled that, for instance, in special
relativity the conservation laws of energy--momentum type are
defined by the global group of automorphisms (the Poincare group)
of the fundamental Mikowski spaces. For (pseudo) Riemannian spaces
one has only tangent space's automorphisms and for particular
cases there are symmetries generated by Killing vectors. No global
or local automorphisms exist on generic la--spaces and in result
of this fact the formulation of la--conservation laws is
sophisticate and full of ambiguities. Nevertheless, a variant of
definition of energy--momentum values for gravitational and matter
la--fields and the form of conservation laws for such values is
possible if we introduce moving frames and consider that
anisotropies are effectively modeled on (pseudo) Riemannian
spacetimes.

The second aim of this paper is to develop a necessary geometric
background (the theory of nearly autoparallel maps, in brief,
na--maps, and tensor integral formalism on multispaces) for
formulation and a detailed investigation of conservation laws on
locally isotropic and/or anisotropic curved spaces. We shall adapt
for the la--spacetimes induced by anholonomic structures
 some results  from the theory of na--maps for generalized
affine spaces, Einstein-Cartan and Einstein spaces, fibre bundles
and different subclasses of generic la--spaces (see
\cite{voa,gv,goz} and \cite {vm} as reviews of our results
published in some less accessible books and journals from former
URSS and Romania).

The question of definition of tensor integration as the inverse
operation of covariant derivation was posed and studied by
A.Mo\'or \cite{moo}. Tensor--integral and bitensor formalisms
turned out to be very useful in solving certain problems connected
with conservation laws in general relativity \cite{goz,vm}. In
order to extend tensor--integral constructions we proposed to take
into consideration nearly autoparallel and nearly geodesic
\cite{sin,vm} maps (in brief, we write ng--maps, ng--theory) which
forms a subclass of local 1--1 maps of curved spaces with
deformation of the connection and metric structures. The third
purpose of this work is to synthesize the results on na--maps and
multispace tensor integrals, to reformulate them for anholonomic
(pseudo) Riemannian spacetimes and to propose a variant of
definition of conservation laws and energy--momentum type values
on la--spacetimes.

Our investigations are completed by some explicit examples of new
solutions of Einstein equations in general relativity which admit
nearly autoparallel maps and/or Finsler like structures.

The paper is organized as follows:\ Section 2 outlines the
geometry of anholonomic frames with associated nonlinear
connection structures. The general criteria when a Finsler like
metric could be embedded into the Einstein gravity is formulated.
Section 3 is devoted to the theory of nearly autoparallel (na)
maps of la--spaceti\-mes. The classification of na--maps and
corresponding invariant conditions are given in Section 4. In
Section 5 we define the nearly autoparallel tensor--integral on
locally anisotropic multispaces. The problem of formulation of
conservation laws on spaces with local anisotropy is studied in
Section 6. We present a definition of conservation laws for
la--gravitational fields on na--images of la--spaces in Section 7.
Some new classes of vacuum and non--vacuume
 solutions  of the Einstein equations, induced by Finsler like metrics, are
constructed in Section 8. In Section 9 we illustrate how a class
of vacuum Einstein fields
 with Finsler like structures can be mapped via  chains of na--transforms
  to the flat Minkowski spacetime. The results are outlined in Section 10.

\section{ Anholonomic Frames and Anisotropic Metrics}

We outline the geometric background of the Einstein gravity with
respect to anholonomic frames and associated nonlinear connections
(N--connection) modelling $m$ dimensional local anisotropies in
(pseido) Riemannian $\left( n+m\right) $--dimensional spacetimes.
Comparing with another approaches concerning anholonomic frames
(tetrads or vierbiends, in four dimensions) in general relativity
(see, for instance, \cite{ellis,mtw,mielke}), it should be noted
that topic of application of N--connections in general relativity
and Kaluza--Klein theories was not considered in the well known
monographs and works. The N--connection geometry was investigated
in details, from the viewpoint of modelling of generalized
Lagrange and Finsler geometries in vector and tangent bundle
spaces,  in the monographs \cite{ma87,ma94,bej} and, with
applications to locally anisotropic (la) spinor differential
geometry,
 supergravity and superstrings in \cite{vjmp,vjhep,vap,vnp}. We pointed
to the positive necessity to consider anholonomic frames with
associated N--connections (with mixed holonomic and anholonomic
degrees of friedom) in order to develop self--consistent
relativistic theories with generic anisotropy, anisotropic
distributions of matter and field interactions, kinetic and
thermodynamic processes on (pseudo)\ Riemannian spacetimes in
\cite{v2000}. The final purpose of this Section is to  proof that
la--structures (Finsler like or more general ones) could be
induced in general relativity.

\subsection{Anholonomy, local anisotropy, and Einstein equations}

In this work spacetimes are modeled as smooth (i. e. class $C^\infty )$ $%
\left( n+m\right) $--dimen\-si\-o\-nal (pseudo) Riemannian
manifolds $V^{(n+m)}$ being Hausdorff, paracompact and connected.
A spacetime $V^{(n+m)}$ is enabled with the fundamental structures
of symmetric metric $g_{\alpha \beta }$ and of linear, in general
nonsymmetric (if we consider anholonomic frames), connection
$\Gamma _{~\beta \gamma }^\alpha $ defining the covariant
derivation $D_\alpha $ which is chosen to satisfy the metricity
conditions $D_\alpha g_{\beta \gamma }=0.$ The indices of
geometrical objects are given with respect to a frame vector field
$\delta ^\alpha =(\delta ^i,\delta ^a)$ and its dual $\delta
_\alpha =(\delta _i,\delta _a).$ For instance, a
covariant--contravariant tensor $Q$ is decomposed as $$ Q=Q_\alpha
^{~\beta }\delta ^\alpha \otimes \delta _\beta , $$
where $\otimes $ is the tensor product. A holonomic frame structure on $%
V^{(n+m)}$ could be stated by a local coordinate base
\begin{equation}
\label{pder}\partial _\alpha =\partial /\partial u^\alpha ,
\end{equation}
consisting from usual partial derivative components, and the dual
basis
\begin{equation}
\label{pdif}d^\alpha =du^\alpha ,
\end{equation}
consisting from usual differentials. An arbitrary holonomic frame
$e_\alpha$
 could be related to a coordinate one by a local linear transform
$ e_\alpha =A_\alpha ^{~\beta }\partial _\beta, $ for which
 the matrix $A_\alpha ^{~\beta }$ is nondegenerate and there are
satisfied the holonomy conditions, $e_\alpha e_\beta -e_\beta
e_\alpha =0. $

Let us consider a $\left( n+m\right) $--dimensional metric
parametrized
\begin{equation}
\label{ansatz}g_{\alpha \beta }=\left[
\begin{array}{cc}
g_{ij}+N_i^aN_j^bh_{ab} & N_j^eh_{ae} \\ N_i^eh_{be} & h_{ab}
\end{array}
\right]
\end{equation}
with respect to a local coordinate basis (\ref{pdif}), $du^\alpha
=\left( dx^i,dy^a\right) ,$ where the Greek indices run values
$1,2,...,n+m,$ the
Latin indices $i,j,k,...$ from the middle of the alphabet run values $%
1,2,...,n$ and the Latin indices from the beginning of the
alphabet, $a,b,c,...,\,$ run values $1,2,...,m.$ The coefficients
$g_{ij}=g_{ij}\left( u^\alpha \right) ,h_{ae}=h_{ae}\left(
u^\alpha \right) $ and $N_i^a=N_i^a(u^\alpha )$ will be defined by
a solution of the Einstein gravitational field equations.

The metric (\ref{ansatz}) can be rewritten in a block $(n\times
n)\oplus (m\times m)$ form
\begin{equation}
\label{dm}g_{\alpha \beta }=\left(
\begin{array}{cc}
g_{ij}(u^\alpha ) & 0 \\ 0 & h_{ab}(u^\alpha )
\end{array}
\right)
\end{equation}
with respect to the anholonomic bases
\begin{equation}
\label{dder}\delta _\alpha =(\delta _i,\partial _a)=\frac \delta
{\partial u^\alpha }=\left( \delta _i=\frac \delta {\partial
x^i}=\frac \partial {\partial x^i}-N_i^b\left( u^\alpha \right)
\frac \partial {\partial y^b},\partial _a=\frac \partial {\partial
y^a}\right)
\end{equation}
and
\begin{equation}
\label{ddif}\delta ^\beta =\left( d^i,\delta ^a\right) =\delta
u^\beta =\left( d^i=dx^i,\delta ^a=\delta y^a=dy^a+N_k^a\left(
u^\alpha \right) dx^k\right) .
\end{equation}
where the coefficients $N_j^a\left( u^\alpha \right) $ from
(\ref{dder}) and (\ref{ddif}) are treated as the components of an
associated nonlinear connection, N--connection, structure
\cite{barth,ma87,ma94,vjmp,vjhep,vnp}.

A frame (local basis) structure $\delta _\alpha $ on $V^{(n+m)}$
is characterized by its anholonomy coefficients $w_{~\beta \gamma
}^\alpha $ defined from relations
\begin{equation}
\label{anholon}\delta _\alpha \delta _\beta -\delta _\beta \delta
_\alpha =w_{~\alpha \beta }^\gamma \delta _\gamma .
\end{equation}

The rigorous mathematical definition of N--connection is based on
the formalism of horizontal and vertical subbundles and on exact
sequences in vector bundles \cite{barth,ma94}. In this work we
introduce a N--connection as a distribution which for every point
$u=(x,y)\in V^{(n+m)}$ defines a local decomposition of the
tangent space $$ T_uV^{(n+m)}=H_uV^{(n)}\oplus V_uV^{(m)}. $$ into
horizontal, $H_uV^{(n)},$ and vertical (anisotropy), $V_uV^{(m)},$
subspaces which is given by a set of coefficients $N_j^a\left(
u^\alpha \right) .$

A N--connection is characterized by its curvature
\begin{equation}
\label{ncurv}\Omega _{ij}^a=\frac{\partial N_i^a}{\partial x^j}-\frac{%
\partial N_j^a}{\partial x^i}+N_i^b\frac{\partial N_j^a}{\partial y^b}-N_j^b%
\frac{\partial N_i^a}{\partial y^b}.
\end{equation}
Here we note that the class of usual linear connections can be
considered as a particular case of N--connections when $$
N_j^a(x,y)=\Gamma _{bj}^a(x)y^b. $$

The elongation (by N--connection) of partial derivatives and
differentials in the adapted to the N--connection operators
(\ref{dder}) and (\ref{ddif}) reflects the fact that on the
(pseudo) Riemannian spacetime $V^{(n+m)}$ it is modeled a generic
local anisotropy characterized by anholonomy relations
(\ref{anholon}) when the anolonomy coefficients are computed as
follows
\begin{eqnarray}
w_{~ij}^k & = & 0,w_{~aj}^k=0,w_{~ia}^k=0,w_{~ab}^k=0,w_{~ab}^c=0,
\nonumber\\ w_{~ij}^a & = & -\Omega _{ij}^a,w_{~aj}^b=-\partial
_aN_i^b,w_{~ia}^b=\partial _aN_i^b. \nonumber
\end{eqnarray}
The frames (\ref{dder}) and (\ref{ddif}) are locally adapted to
the N--connection structure, define a local anisotropy and, in
brief, are called la--bases.

A N--connection structure distinguishes the geometrical objects
into horizontal and vertical components. Such objects are briefly
called d--tensors, d--metrics and d--connections. Their components
are defined with respect to a la--basis of type (\ref{dder}), its
dual (\ref{ddif}), or their tensor products (d--linear or
d--affine transforms of such frames could also be considered). For
instance, a covariant and contravariant d--tensor $Q,$ is
expressed $$ Q=Q_{~\beta }^\alpha \delta _\alpha \otimes \delta
^\beta =Q_{~j}^i\delta _i\otimes d^j+Q_{~a}^i\delta _i\otimes
\delta ^a+Q_{~j}^b\partial _b\otimes d^j+Q_{~a}^b\partial
_b\otimes \delta ^a. $$

In this paper, as a {\bf locally anisotropic spacetime,
la--spacetime}, we shall consider a pseudo--Riemannian spacetime
provided with a metric of signature $(-, +,$ $ +, +)$ (a
permutation of signes being also possible) and with an anholonomic
frame basis defined by an  associated N--connection structure when
the coefficients of the mentioned objects are imposed to
 be a solution of the Einstein equations.

A linear d--connection $D$ on la--spacetime $V^{(n+m)}{\cal ,}$ $$
D_{\delta _\gamma }\delta _\beta =\Gamma _{~\beta \gamma }^\alpha
\left( x,y\right) \delta _\alpha , $$ is given by its
h--v--components,
\begin{equation}
\label{dcon}\Gamma _{~\beta \gamma }^\alpha =\left(
L_{~jk}^i,L_{~bk}^a,C_{~jc}^i,C_{~bc}^a\right)
\end{equation}
where%
$$ D_{\delta _k}\delta _j=L_{~jk}^i\delta _i,D_{\delta _k}\partial
_b=L_{bk}^a\partial _a,D_{\partial _c}\delta _j=C_{~jc}^i\delta
_i,D_{\delta _c}\partial _b=C_{bc}^a\partial _a. $$

A metric  on $V^{(n+m)}$ with its coefficients
 parametrized as (\ref{ansatz}) can be written in
distingushed form (\ref{dm}), as a metric d--tensor (in brief,
d--metric), with respect to  a la--base (\ref{ddif}), i. e.
\begin{equation}
\label{dmetric}\delta s^2=g_{\alpha \beta }\left( u\right) \delta
^\alpha \otimes \delta ^\beta
=g_{ij}(x,y)dx^idx^j+h_{ab}(x,y)\delta y^a\delta y^b.
\end{equation}

Some N--connection, d--connection and d--metric structures are
compatible if there are satisfied the conditions $$ D_\alpha
g_{\beta \gamma }=0. $$ For instance, a canonical compatible
d--connection $$ ^c\Gamma _{~\beta \gamma }^\alpha =\left(
^cL_{~jk}^i,^cL_{~bk}^a,^cC_{~jc}^i,^cC_{~bc}^a\right) $$ is
defined by the coefficients of d--metric (\ref{dmetric}),
$g_{ij}\left( x,y\right) $ and $h_{ab}\left( x,y\right) ,$ and by
the coefficients of
N--connection,%
\begin{eqnarray}
^cL_{~jk}^i & = & \frac 12g^{in}\left( \delta _kg_{nj}+\delta
_jg_{nk}-\delta _ng_{jk}\right) , \label{cdcon} \\ ^cL_{~bk}^a & =
& \partial _bN_k^a+\frac 12h^{ac}\left( \delta
_kh_{bc}-h_{dc}\partial _bN_i^d-h_{db}\partial _cN_i^d\right) ,
\nonumber \\ ^cC_{~jc}^i & = & \frac 12g^{ik}\partial _cg_{jk},
\nonumber \\ ^cC_{~bc}^a & = & \frac 12h^{ad}\left( \partial
_ch_{db}+\partial _bh_{dc}-\partial _dh_{bc}\right)  \nonumber
\end{eqnarray}
The coefficients of the canonical d--connection generalize for
la--spacetimes the well known Cristoffel symbols. By a local
linear nondegenerate transform to a coordinate frame we obtain the
coefficients of the usual (pseudo) Riemannian metric connection.

For a canonical d--connection (\ref{dcon}) the components of
canonical torsion,
\begin{eqnarray}
&T\left( \delta _\gamma ,\delta _\beta \right) &=T_{~\beta \gamma
}^\alpha \delta _\alpha ,  \nonumber \\ &T_{~\beta \gamma }^\alpha
&= \Gamma _{~\beta \gamma }^\alpha -\Gamma _{~\gamma \beta
}^\alpha +w_{~\beta \gamma }^\alpha \nonumber
\end{eqnarray}
are expressed via d--torsions
\begin{eqnarray}
T_{.jk}^i & = & T_{jk}^i=L_{jk}^i-L_{kj}^i,\quad
T_{ja}^i=C_{.ja}^i,T_{aj}^i=-C_{ja}^i, \nonumber \\ T_{.ja}^i & =
& 0,\quad T_{.bc}^a=S_{.bc}^a=C_{bc}^a-C_{cb}^a, \label{dtorsions}
\\ T_{.ij}^a & = & -\Omega _{ij}^a,\quad T_{.bi}^a= \partial _b
N_i^a -L_{.bj}^a,\quad T_{.ib}^a=-T_{.bi}^a \nonumber
\end{eqnarray}
which reflects the anholonomy of the corresponding
 la--frame of reference on $V^{(n+m)};$ they
are induced effectively. With respect to holonomic frames the
d--torsions vanishes.

For simplicity, hereafter,   we shall omit the up left index ''c''
and consider only connections and d--connections defined by
compatible metric and N--connection coefficients.

Putting the non--vanishing coefficients (\ref{dcon}) into the
formula for curvature
\begin{eqnarray}
R\left( \delta _\tau ,\delta _\gamma \right) \delta _\beta &= &
R_{\beta ~\gamma\tau }^{~\alpha }\delta _\alpha , \nonumber \\
R_{\beta ~\gamma \tau }^{~\alpha } & = & \delta _\tau \Gamma
_{~\beta \gamma }^\alpha -\delta _\gamma \Gamma _{~\beta \delta
}^\alpha +
 \Gamma _{~\beta \gamma }^\varphi \Gamma _{~\varphi \tau }^\alpha
-\Gamma _{~\beta \tau }^\varphi \Gamma _{~\varphi \gamma }^\alpha
+ \Gamma _{~\beta \varphi }^\alpha w_{~\gamma \tau }^\varphi
\nonumber
\end{eqnarray}
we compute the components of canonical d--curvatures
\begin{eqnarray}
R_{h.jk}^{.i} & = & \delta _kL_{.hj}^i-\delta_jL_{.hk}^i
 +  L_{.hj}^mL_{mk}^i-L_{.hk}^mL_{mj}^i-C_{.ha}^i\Omega _{.jk}^a,
\nonumber \\ R_{b.jk}^{.a} & = & \delta
_kL_{.bj}^a-\delta_jL_{.bk}^a
 +  L_{.bj}^cL_{.ck}^a-L_{.bk}^cL_{.cj}^a-C_{.bc}^a\Omega _{.jk}^c,
\nonumber \\ P_{j.ka}^{.i} & = & \partial _kL_{.jk}^i
+C_{.jb}^iT_{.ka}^b
 -  ( \partial _kC_{.ja}^i+L_{.lk}^iC_{.ja}^l -
L_{.jk}^lC_{.la}^i-L_{.ak}^cC_{.jc}^i ), \nonumber \\
P_{b.ka}^{.c} & = & \partial _aL_{.bk}^c +C_{.bd}^cT_{.ka}^d
  - ( \partial _kC_{.ba}^c+L_{.dk}^{c\,}C_{.ba}^d
- L_{.bk}^dC_{.da}^c-L_{.ak}^dC_{.bd}^c ), \nonumber \\
S_{j.bc}^{.i} & = & \partial _cC_{.jb}^i-\partial _bC_{.jc}^i
 +  C_{.jb}^hC_{.hc}^i-C_{.jc}^hC_{hb}^i, \nonumber \\
S_{b.cd}^{.a} & = &\partial _dC_{.bc}^a-\partial
_cC_{.bd}^a+C_{.bc}^eC_{.ed}^a-C_{.bd}^eC_{.ec}^a. \nonumber
\end{eqnarray}

The Ricci d--tensor $R_{\beta \gamma }=R_{\beta ~\gamma \alpha
}^{~\alpha }$ has the components
\begin{eqnarray}
R_{ij} & = & R_{i.jk}^{.k},\quad
 R_{ia}=-^2P_{ia}=-P_{i.ka}^{.k},\label{dricci} \\
R_{ai} &= & ^1P_{ai}=P_{a.ib}^{.b},\quad R_{ab}=S_{a.bc}^{.c}
\nonumber
\end{eqnarray}
and, in general, this d--tensor is non symmetric.

We can compute the scalar curvature $\overleftarrow{R}=g^{\beta
\gamma
}R_{\beta \gamma }$ of a d-connection $D,$%
\begin{equation}
\label{dscalar}{\overleftarrow{R}}=\widehat{R}+S,
\end{equation}
where $\widehat{R}=g^{ij}R_{ij}$ and $S=h^{ab}S_{ab}.$

By introducing the values (\ref{dricci}) and (\ref{dscalar}) into
the usual Einstein equations $$ R_{\beta \gamma }-\frac 12g_{\beta
\gamma }R=k\Upsilon _{\beta \gamma }, $$ written with respect to
an anholonomic la--frame of reference, we obtain the system of
field equations for la--gravity with N--connection structure \cite
{ma94}:%
\begin{eqnarray}
R_{ij}-\frac 12\left( \widehat{R}+S\right) g_{ij} & = & k\Upsilon
_{ij}, \label{einsteq2} \\ S_{ab}-\frac 12\left(
\widehat{R}+S\right) h_{ab} & = & k\Upsilon _{ab},
 \nonumber \\
^1P_{ai} & = & k\Upsilon _{ai}, \nonumber \\ ^2P_{ia} & = &
-k\Upsilon _{ia}, \nonumber
\end{eqnarray}
where $\Upsilon _{ij},\Upsilon _{ab},\Upsilon _{ai}$ and $\Upsilon
_{ia}$ are the components of the energy--momentum d--tensor field
$\Upsilon _{\beta \gamma }$ which includes the cosmological
constant terms and possible contributions of d--torsions and
matter, and $k$ is the coupling constant.

\subsection{Finsler like metrics in Einstein gravity}

In this subsection we follow the almost Hermitian model of Finsler
geometry \cite{ma87,ma94} and consider a $V^{2n}$ (pseudo)
Riemannian spacetime.

The locally anisotropic structure is modeled on the manifold $\widetilde{TV}%
=TV^{(n)}\backslash \{0\},$ where $\backslash \{0\}$ means that
there is eliminated the null cross--section of the bundle
projection $\tau :TV^{(n)}\rightarrow V^{(n)}).$ There are
considered d--metrics of type (\ref {dmetric}) with identical
$(n\times n)$--dimensional blocks for both base and fiber
components. On $TV^{(n)}$ we can define a natural almost complex
structure $C_{(a)},$ $$
C_{(a)}\left( \delta _i\right) =-\partial /\partial y^i\mbox{ and }%
C_{(a)}\left( \partial /\partial y^i\right) =\delta _i, $$ where
the la--derivatives (\ref{dder}) and la--differentials act on the
bundle  $\widetilde{TV}$ being adapted to a nontrivial N--connection $%
N=\{N_j^k\left( x,y\right) \}$ in $TV$ and $C_{(a)}^2=-I.$ The
pair $\left(
\delta s^2,C_{(a)}\right) $ defines an almost Hermitian structure on $%
\widetilde{TV}$ with an associate 2--form%
$$ \theta =h_{ij}\left( x,y\right) \delta ^i\wedge dx^j $$ and the
triad $K^{2n}=\left( \widetilde{TV},\delta s^2,C_{(a)}\right) $ is
an almost K\"ah\-le\-ri\-an space. We can verify that the
canonical d--connection (\ref{dcon}) satisfies the conditions $$
~^cD_X\left( \delta s^2\right) =0,~^cD_X\left( C_{(a)}\right) =0
$$ for any d--vector $X$ on $TV^{(n)}$ and has zero $hhh$-- and
$vvv$--torsions (where h-- and v-- denote the horizontal and
vertical components).

The notion of {\bf Lagrange space} \cite{kern,ma87,ma94} was
introduced as a generalization of Finsler geometry in order to
geometrize the fundamental
concepts in mechanics. A regular Lagrangian $L\left( x^i,y^i\right) $ on $%
\widetilde{TV}$ is defined by a continuity class $C^\infty $ function $%
L:TV^{(n)}\rightarrow \R$ for which the matrix
\begin{equation}
\label{lagrm}h_{ij}\left( x,y\right) =\frac 12\frac{\partial
^2L}{\partial y^i\partial y^j}
\end{equation}
has the rank $n.$ A d--metric (\ref{dmetric}) with coefficients of
form (\ref {lagrm}), a corresponding canonical d--connection
(\ref{dcon}) and almost complex structure $C_{(a)}$ defines an
almost Hermitian model of Lagrange geometry.

Metrics $h_{ij}\left( x,y\right) $ of rank $n$ and constant signature on $%
\widetilde{TV}$, which can not be determined as a second
derivative of a Lagrangian are considered in the so--called
generalized Lagrange geometry on $TV^{(n)}$ (see details in
\cite{ma87,ma94}).

A subclass of metrics of type (\ref{lagrm}) consists from those
where instead of a regular Lagrangian one considers a Finsler
metric function $F$ on $V^{(n)}$ defined as $F:TV^{(n)}\rightarrow
\R$ having the properties that it is of class $C^\infty $ on
$\widetilde{TV}$ and only continuous on the image of the null
cross--section in $TV^{(n)},$ the restriction of $F$ on
$\widetilde{TV}$ is a positive function homogeneous of degree 1
with respect to the variables $y^i,$ i. e. $$ F\left( x,\lambda
y\right) =\lambda F\left( x,y\right) ,\lambda \in \R, $$ and the
quadratic form on $\R^2,$ with coefficients
\begin{equation}
\label{finm2}h_{ij}\left( x,y\right) =\frac 12\partial
^2F^2/\partial y^i\partial y^j
\end{equation}
(see (\ref{dmetric})) given on $\widetilde{TM,}$ is positive
definite.

Very different approaches to Finsler geometry, its generalizations
and applications are examined in a number of monographs \cite
{fin,car35,rund,as,asp,bej,bog2,ma87} and it is considered that
for such geometries the usual (pseudor)Riemannian metric interval
$$ ds=\sqrt{g_{ij}\left( x\right) dx^idx^j} $$ on a manifold $M$
is changed into a nonlinear one
\begin{equation}
\label{finint}ds=F\left( x^i,dx^j\right)
\end{equation}
defined by the Finsler metric $F$ (fundamental function) on
$\widetilde{TM}$ (it should be noted an ambiguity in terminology
used in monographs on Finsler geometry and on gravity theories
with respect to such terms as Minkowski space, metric function and
so on).

Geometric spaces with a 'combersome' variational calculus and a
number of curvatures, torsions and invariants connected with
nonlinear metric intervals of type (\ref{lagrm}) are considered as
less suitable for purposes of the modern field and particle
physics.

In our approach to generalized Finsler geometries in (super)
string, gravity and gauge theories \cite{vnp,vm} we advocated the
idea that instead of usual geometric constructions based on
straightforward applications of derivatives of (\ref{finm2})
following from a nonlinear interval (\ref{finint}) one should
consider d--metrics (\ref{dmetric}) with coefficients of necessity
determined via an almost Hermitian model of a Lagrange
(\ref{lagrm}), Finsler geometry (\ref{finm2}) and/or their
extended variants. This way, by a synthesis of the moving frame
method with the geometry of N--connections, we can investigate in
a unified manner a various class of higher and lower dimension
gravitational models with generic, or induced, anisotropies on
some anholonomic and/or Kaluza--Klein spacetimes.

Now we analyze the possibility to include $n$--dimensional Finsler
metrics into $2n$--dimensional (pseudo) Riemannian spaces and
formulate the general criteria when a Finsler like metric could be
imbedded into the Einstein theory.

Let consider on $\widetilde{TV}$ an ansatz of type (\ref{ansatz})
when $$
g_{ij}=\frac 12\partial ^2F_{*}^2/\partial y^i\partial y^j\mbox{ and }%
h_{ij}=\frac 12\partial ^2F^2/\partial y^i\partial y^j $$ i.e.
\begin{equation}
\label{ffm}g_{\widehat{\alpha }\widehat{\beta }}=\frac 12\left(
\begin{array}{cc}
\frac{\partial ^2F_{*}^2}{\partial y^i\partial
y^j}+N_i^kN_j^l\frac{\partial ^2F^2}{\partial y^k\partial y^l} &
N_j^l
\frac{\partial ^2F^2}{\partial y^k\partial y^l} \\ N_i^k\frac{\partial ^2F^2%
}{\partial y^k\partial y^l} & \frac{\partial ^2F^2}{\partial
y^i\partial y^j}
\end{array}
\right) .
\end{equation}
A metric $g_{\widehat{\alpha }\widehat{\beta }}$ of signature
$(-,+,...,+)$ induced by two Finsler functions $F_{*}$ and $F$
(\ref{finm2}) (as a particular
 case  $F_{*}=F)$ is to be treated in the framework of general
relativity theory if this metric is a solution of the Einstein
equations on a $2n$--dimensional (pseudo) Riemannian spacetime
written with respect to a holonomic frame. Here we note that, in
general, a N--connection on a Finsler space, subjected to the
condition that the induced (pseudo) Riemannian metric is a
solution of usual Einstein equations, does not coincide with the
well known Cartan's N--connection in Finsler geometry
\cite{car35,rund}. In such cases we have to examine possible
compatible deformations of N--connection structures
\cite{ma87,ma94}.

We can also introduces ansatzs of type (\ref{ansatz}) with $g_{ij}$ and $%
h_{ij} $ induced by a Lagrange quadratic form (\ref{lagrm}). In
Section 8
 we shall construct solutions of the Einstein equations forllowing from
 an ansatz for a generalized Finsler metric,
\begin{equation}
\label{glm}g_{\widehat{\alpha }\widehat{\beta }}=\frac 12\left(
\begin{array}{cc}
\frac{\partial ^2F_{*}^2}{\partial y^i\partial
y^j}+N_i^kN_j^l{\Lambda}_{ij} & N_j^l
 {\Lambda}_{kl}\\ N_i^k{\Lambda}_{kl} &
 {\Lambda}_{ij}
\end{array}
\right),
\end{equation}
where the $2\times 2$ matrix $\Lambda$ is induced by a Finsler
metric via a transform
\begin{equation} \label{diagonalization}
{\Lambda}_{kl}=(C^T)_{ki}\frac{\partial ^2F^2}{\partial
y^i\partial y^j}(C)_{jl},
\end{equation}
parametrized by a $2\times 2$ matrix $(C)(x^i,y^k)$  and its
transposition  $(C^T)(x^i,y^k).$ A general approach to the
geometry of spacetimes with generic local anisotropy can be
developed on imbeddings into corresponding Kaluza--Klein theories
and adequate modelling of la--interactions with respect to
anholonomic or holonomic frames and associated N--connection
structures. As a matter of principle every type of Finsler,
Lagrange or generalized Lagrange  geometry could be modeled on a
corresponding Kaluza--Klein spacetime.

\section{Nearly Autoparallel Maps}

The aim of this Section is to formulate the theory of nearly
geodesic maps (ng--maps) \cite{sin} and nearly autoparallel maps
(na--maps) \cite{vm,voa} for (pseudo) Riemannian spacetimes
provided with anhonlonomic frame and associated N--connection
structures.

Our geometric arena consists from pairs of open regions
$(U,{\underline{U}})$ of two la--space\-ti\-mes, $U{\subset
}V^{(n+m)}$ and ${\underline{U}}{\subset
}\underline{{V}}{^{(n+m)}}$, and 1--1 local maps $f:U{\to
}{\underline{U}}$ given by some functions $f^\alpha (u)$ of
smoothly class $C^r{(U)},(r>2,$ or
$r={\omega }$~ for analytic functions) and their inverse functions $f^{%
\underline{\alpha }}({\underline{u}})$ with corresponding
non--zero
Jacobians in every point $u{\in }U$ and ${\underline{u}}{\in }{\underline{U}}%
.$

We consider that two open regions $U$~ and ${\underline{U}}$~ are
attributed to a common for f--map coordinate system if this map is
realized on the principle of coordinate equality $q(u^\alpha ){\to
}{\underline{q}}(u^\alpha
)$~ for every point $q{\in }U$~ and its f--image ${\underline{q}}{\in }{%
\underline{U}}.$ We note that all calculations included in this
work will be
local in nature and taken to refer to open subsets of mappings of type $%
V^{(n+m)}{\supset }U\stackrel{f}{\longrightarrow }{\underline{U}}{\subset }%
\underline{{V}}{^{(n+m)}}.$ %
For simplicity, we suppose that in a fixed common coordinate
system for $U$ and ${\underline{U}}$ the spacetimes $V^{(n+m)}$
and $\underline{{V}}{^{(n+m)}}$ are characterized by a common
N--connection structure, when $$
N_j^a(u)={\underline{N}}_j^a(u)={\underline{N}}_j^a({\underline{u}}),
$$ which leads to the possibility to establish common local bases,
adapted to a given N--connection, on both regions $U$ and
${\underline{U}.}$ We consider that on $V^{(n+m)}$ it is defined a
linear d--connection structure with
components ${\Gamma }_{{.}{\beta }{\gamma }}^\alpha .$ On the space $%
\underline{{V}}{^{(n+m)}}$ the linear d--connection is considered
to be a general one with torsion $$
{\underline{T}}_{{.}{\beta }{\gamma }}^\alpha ={\underline{\Gamma }}_{{.}{%
\beta }{\gamma }}^\alpha -{\underline{\Gamma }}_{{.}{\gamma }{\beta }%
}^\alpha +w_{{.}{\beta }{\gamma }}^\alpha $$ and nonmetricity $$
{\underline{K}}_{{\alpha }{\beta }{\gamma }}={{\underline{D}}_\alpha }{%
\underline{g}}_{{\beta }{\gamma }}. $$ As a particular case we can
consider maps to (pseudo) Riemannian spacetimes, when
${\underline{K}}_{{\alpha }{\beta }{\gamma }}=0.$

Geometrical objects on $\underline{{V}}{^{(n+m)}}$ are specified
by underlined
symbols (for example, ${\underline{A}}^\alpha ,{\underline{B}}^{{\alpha }{%
\beta }})$~ or by underlined indices (for example, $A^{\underline{a}},B^{{%
\underline{a}}{\underline{b}}}).$

For our purposes it is convenient to introduce auxiliary
sym\-met\-ric
d--con\-nec\-ti\-ons, ${\gamma }_{{.}{\beta }{\gamma }}^\alpha ={\gamma }_{{.%
}{\gamma }{\beta }}^\alpha $~ on $V^{(n+m)}$ and ${\underline{\gamma }}_{.{%
\beta }{\gamma }}^\alpha ={\underline{\gamma }}_{{.}{\gamma }{\beta }%
}^\alpha $ on $\underline{{V}}{^{(n+m)}}$ defined,
correspondingly, as $$
{\Gamma }_{{.}{\beta }{\gamma }}^\alpha ={\gamma }_{{.}{\beta }{\gamma }%
}^\alpha +T_{{.}{\beta }{\gamma }}^\alpha {\mbox{ and }\underline{\Gamma }}_{%
{.}{\beta }{\gamma }}^\alpha ={\underline{\gamma }}_{{.}{\beta }{\gamma }%
}^\alpha +{\underline{T}}_{{.}{\beta }{\gamma }}^\alpha . $$

We are interested in definition of local 1--1 maps from $U$ to ${\underline{U%
}}$ characterized by symmetric, $P_{{.}{\beta }{\gamma }}^\alpha
,$ and antisymmetric, $Q_{{.}{\beta }{\gamma }}^\alpha $,~
deformations:
\begin{equation}
\label{simdef}{\underline{\gamma }}_{{.}{\beta }{\gamma }}^\alpha ={\gamma }%
_{{.}{\beta }{\gamma }}^\alpha +P_{{.}{\beta }{\gamma }}^\alpha
\end{equation}
and
\begin{equation}
\label{torsdef}{\underline{T}}_{{.}{\beta }{\gamma }}^\alpha =T_{{.}{\beta }{%
\gamma }}^\alpha +Q_{{.}{\beta }{\gamma }}^\alpha .
\end{equation}
The auxiliary linear covariant derivations induced by ${\gamma }_{{.}{\beta }%
{\gamma }}^\alpha $ and ${\underline{\gamma }}_{{.}{\beta }{\gamma
}}^\alpha
$~ are denoted respectively as $^{({\gamma })}D$~ and $^{({\gamma })}{%
\underline{D}}.$~

Curves on $U$ are parametrized $$
u^\alpha =u^\alpha ({\eta })=(x^i({\eta }),y^i({\eta })),~{\eta }_1<{\eta }<{%
\eta }_2, $$
where the corresponding tangent vector fields are defined%
$$
v^\alpha ={\frac{{du^\alpha }}{d{\eta }}}=({\frac{{dx^i({\eta })}}{{d{\eta }}%
}},{\frac{{dy^j({\eta })}}{d{\eta }}}). $$

\begin{definition}
A curve $l$~ is called auto parallel, a--parallel, on $V^{(n+m)}$
if its tangent vector field $v^\alpha $~ satisfies the a--parallel
equations
\begin{equation}
\label{apareq}v\ Dv^\alpha =v^\beta \ {^{({\gamma })}D}_\beta v^\alpha ={%
\rho }({\eta })v^\alpha ,
\end{equation}
where ${\rho }({\eta })$~ is a scalar function on $V^{(n+m)}$.
\end{definition}

Let a curve ${\underline{l}}{\subset }{\underline{\xi }}$ be given
in parametric form $u^\alpha =u^\alpha ({\eta }),~{\eta }_1<{\eta
}<{\eta }_2$
with the tangent vector field $v^\alpha ={\frac{{du^\alpha }}{{d{\eta }}}}{%
\ne }0.$ We suppose that a 2--dimensional distribution
$E_2({\underline{l}})$
is defined along ${\underline{l}},$ i.e. in every point $u{\in }{\underline{l%
}}$ a 2-dimensional vector space $E_2({\underline{l}}){\subset }{\underline{%
\xi }}$ is fixed. The introduced distribution
$E_2({\underline{l}})$~ is coplanar along ${\underline{l}}$~ if
every vector ${\underline{p}}^\alpha
(u_{(0)}^b){\subset }E_2({\underline{l}}),$ $u_{(0)}^\beta {\subset }{%
\underline{l}}$~ rests contained in the same distribution after
parallel
transports along ${\underline{l}},$~ i.e. ${\underline{p}}^\alpha (u^\beta ({%
\eta })){\subset }E_2({\underline{l}}).$

\begin{definition}
A curve ${\underline{l}}$~ is called nearly autoparallel, or, in
brief, na--parallel, on the spacetime $\underline{{V}}{^{(n+m)}}$~
if a coplanar along ${\underline{l}}$~ distribution
$E_2({\underline{l}})$ containing the
tangent to ${\underline{l}}$~ vector field $v^\alpha ({\eta })$,~ i.e. $%
v^\alpha ({\eta }){\subset }E_2({\underline{l}}),$~ is defined.
\end{definition}

We can define nearly autoparallel maps of la--spacetimes as an
anisotropic generalization of the constructions for the locally
isotropic spaces (see  ng--\cite{sin} and na--maps\cite{voa,vm}):
\begin{definition}
Nearly autoparallel maps, na--maps, of la--spacetimes are defined
as local 1--1 mappings $V^{(n+m)}{\to }\underline{{V}}{^{(n+m)}}$
which change every a--parallel on $V^{(n)}$ into a na--parallel on
$\underline{{V}}{^{(n+m)}}.$
\end{definition}

Now we formulate the general conditions when some deformations (\ref{simdef}%
) and (\ref{torsdef}) charac\-ter\-ize na-maps:

 Let an a-parallel $l{\subset
}U$~ is given by some func\-ti\-ons $u^\alpha =u^{({\alpha })}({\eta }),v^\alpha =%
{\frac{{du^\alpha }}{d{\eta }}}$, ${\eta }_1<{\eta }<{\eta }_2$,
satisfying
 the equations (\ref{apareq}). We suppose that to this a--parallel corresponds a
na--parallel ${\underline{l}}\subset {\underline{U}}$ given by the
same parameterization in a common for a chosen na--map coordinate
system on $U$~
and ${\underline{U}}.$ This condition holds for the vectors ${\underline{v}}%
_{(1)}^\alpha =v{\underline{D}}v^\alpha $~ and $v_{(2)}^\alpha =v\ {%
\underline{D}}v_{(1)}^\alpha $ satisfying the equality
\begin{equation}
\label{lindep}{\underline{v}}_{(2)}^\alpha ={\underline{a}}({\eta
})v^\alpha +{\underline{b}}({\eta }){\underline{v}}_{(1)}^\alpha
\end{equation}
for some scalar functions ${\underline{a}}({\eta })$~ and ${\underline{b}}({%
\eta })$~ (see Definitions 2 and 3). Putting the splittings
(\ref{simdef}) and (\ref{torsdef}) into the expressions for
${\underline{v}}_{(1)}^\alpha $ and ${\underline{v}}_{(2)}^\alpha
$ from (\ref{lindep}) we obtain:
\begin{equation}
\label{neq1}v^\beta v^\gamma v^\delta (D_\beta P_{{.}{\gamma }{\delta }%
}^\alpha +P_{{.}{\beta }{\tau }}^\alpha P_{{.}{\gamma }{\delta }}^\tau +Q_{{.%
}{\beta }{\tau }}^\alpha P_{{.}{\gamma }{\delta }}^\tau
)=bv^\gamma v^\delta P_{{.}{\gamma }{\delta }}^\alpha +av^\alpha ,
\end{equation}
where
\begin{equation}
\label{defpar}b({\eta },v)={\underline{b}}-3{\rho },\qquad
\mbox{and}\qquad
a({\eta },v)={\underline{a}}+{\underline{b}}{\rho }-v^b{\partial }_b{\rho }-{%
\rho }^2
\end{equation}
are called the deformation parameters of na--maps.

The algebraic equations for the deformation of torsion $Q_{{.}{\beta }{\tau }%
}^\alpha $ should be written as the compatibility conditions for a
given
nonmetricity tensor ${\underline{K}}_{{\alpha }{\beta }{\gamma }}$~ on $%
\underline{{V}}{^{(n+m)}}$ ( or as the metricity conditions if the
d--connection ${\underline{D}}_\alpha $~ is required to be
metric):
\begin{equation}
\label{metrcond}D_\alpha G_{{\beta }{\gamma }}-P_{{.}{\alpha }({\beta }%
}^\delta G_{{{\gamma })}{\delta }}-{\underline{K}}_{{\alpha
}{\beta }{\gamma }}=Q_{{.}{\alpha }({\beta }}^\delta G_{{\gamma
}){\delta }},
\end{equation}
where $({\quad })$ denotes the symmetrical alternation.

So, we have proved this

\begin{theorem}
The na--maps from a la--spacetime $V^{(n+m)}$ to la--spacetime $\underline{{V}}%
{^{(n+m)}}$~ with a fixed common nonlinear connection structure, $N_j^a(u)={%
\underline{N}}_j^a(u),$ and given d--connections, ${\Gamma }_{{.}{\beta }{%
\gamma }}^\alpha $~ on $V^{(n+m)}$~ and ${\underline{\Gamma }}_{{.}{\beta }{%
\gamma }}^\alpha $~ on ${\underline{{V}}}{^{(n+m)}},$ are locally
parametrized by the solutions of equations (\ref{neq1}) and
(\ref{metrcond}) for every
point $u^\alpha $~ and direction $v^\alpha $~ on $U{\subset }\underline{{V}}{%
^{(n+m)}}.$
\end{theorem}

We call (\ref{neq1}) and (\ref{metrcond}) the basic equations for
na--maps of la--spacetimes. They generalize the corresponding
Sinyukov's equations \cite{sin} which were introduced for
isotropic spaces provided with symmetric affine connection
structure, hold  for generalized Finsler metrics modeled on vector
and tangent bundle spaces and consist a particular case of the
na--maps of (super) vector bundles provided with N--connection
structures \cite{vm,voa}.

\section{Classification of Na--Maps}

Na--maps are classified on possible polynomial parametrizations on
variables $v^\alpha $~ of deformations parameters $a$ and $b,$
see formulas (\ref{neq1}) and (\ref {defpar})).

\begin{theorem}
There are four classes of na--maps characterized by
cor\-res\-pond\-ing deformation parameters and tensors and basic
equations:

\begin{enumerate}
\item  for $na_{(0)}$--maps, ${\pi }_{(0)}$--maps,
$$
P_{{\beta }{\gamma }}^\alpha (u)={\psi }_{{(}{\beta }}{\delta }_{{\gamma }%
)}^\alpha $$
(${\delta }_\beta ^\alpha $~ is Kronecker symbol and ${\psi }_\beta ={\psi }%
_\beta (u)$~ is a covariant vector d--field);

\item  for $na_{(1)}$--maps
$$ a(u,v)=a_{{\alpha }{\beta }}(u)v^\alpha v^\beta ,\quad
b(u,v)=b_\alpha (u)v^\alpha $$ and $P_{{.}{\beta }{\gamma
}}^\alpha (u)$~ is the solution of equations
\begin{equation}
\label{neqc1}D_{({\alpha }}P_{{.}{\beta }{\gamma })}^\delta +P_{({\alpha }{%
\beta }}^\tau P_{{.}{\gamma }){\tau }}^\delta -P_{({\alpha }{\beta
}}^\tau
Q_{{.}{\gamma }){\tau }}^\delta =b_{({\alpha }}P_{{.}{\beta }{\gamma })}^{{%
\delta }}+a_{({\alpha }{\beta }}{\delta }_{{\gamma })}^\delta ;
\end{equation}

\item  for $na_{(2)}$--maps
\begin{eqnarray}
a(u,v) &= &a_\beta (u)v^\beta ,\quad b(u,v)={\frac{{b_{{\alpha }{\beta }}v^\alpha v^\beta }}{{{\sigma }_\alpha (u)v^\alpha }}},\quad {\sigma }%
_\alpha v^\alpha {\neq }0,  \nonumber \\ P_{{.}{\alpha }{\beta
}}^\tau (u) &= & {{\psi }_{({\alpha }}}{\delta }_{{\beta })}^\tau
+{\sigma }_{({\alpha }}F_{{\beta })}^\tau  \label{defcon2}
\end{eqnarray}
and $F_\beta ^\alpha (u)$~ is the solution of equations
\begin{equation}
\label{neqc2}{D}_{({\gamma }}F_{{\beta })}^\alpha +F_\delta ^\alpha F_{({%
\gamma }}^\delta {\sigma }_{{\beta })}-Q_{{.}{\tau }({\beta }}^\alpha F_{{%
\gamma })}^\tau ={\mu }_{({\beta }}F_{{\gamma })}^\alpha +{\nu }_{({\beta }}{%
\delta }_{{\gamma })}^\alpha
\end{equation}
$({\mu }_\beta (u),{\nu }_\beta (u),{\psi }_\alpha (u),{\sigma }_\alpha (u)$%
~ are covariant d--vectors);

\item  for $na_{(3)}$--maps
\begin{eqnarray}
b(u,v) &=&{\frac{{{\alpha }_{{\beta }{\gamma }{\delta }}v^\beta
v^\gamma v^\delta }}{{{\sigma }_{{\alpha }{\beta }}v^\alpha
v^\gamma }}}, \nonumber \\ P_{{.}{\beta }{\gamma }}^\alpha (u) &=&
{\psi }_{({\beta }}{\delta }_{{\gamma })}^\alpha + {\sigma
}_{{\beta }{\gamma }}{\varphi }^\alpha , \label{defcon3}
\end{eqnarray}
where ${\varphi }^\alpha $~ is the solution of equations
\begin{equation}
\label{neqc3}D_\beta {\varphi }^\alpha ={\nu }{\delta }_\beta ^\alpha +{\mu }%
_\beta {\varphi }^\alpha +{\varphi }^\gamma Q_{{.}{\gamma }{\delta
}}^\alpha ,
\end{equation}
${\alpha }_{{\beta }{\gamma }{\delta }}(u),{\sigma }_{{\alpha }{\beta }}(u),{%
\psi }_\beta (u),{\nu }(u)$~ and ${\mu }_\beta (u)$~ are
d--tensors.
\end{enumerate}
\end{theorem}

{\sf Proof.} We sketch the proof respectively for every point in
the theorem:

\begin{enumerate}
\item  It is easy to verify that a--parallel equations (\ref{apareq}) on $%
V^{(n+m)}$ transform into similar ones on
$\underline{{V}}{^{(n+m)}}$ if and
only if deformations (\ref{simdef}) with deformation d--tensors of type ${%
P^\alpha }_{\beta \gamma }(u)={\psi }_{(\beta }{\delta }_{\gamma
)}^\alpha $ are considered.

\item  Using corresponding to $na_{(1)}$--maps parametrizations of $a(u,v)$
and $b(u,v)$ (see conditions of the theorem) for arbitrary
$v^\alpha \neq 0$ on $U\in V^{(n+m)}$ and after a redefinition of
deformation parameters we obtain that equations (\ref{neqc1}) hold
if and only if ${P^\alpha }_{\beta \gamma }$ satisfies
(\ref{simdef}).

\item  In a similar manner we obtain basic $na_{(2)}$--map equations (\ref
{neqc2}) from (\ref{neq1}) by considering
$na_{(2)}$--parametrizations of deformation parameters and
d--tensor.

\item  For $na_{(3)}$--maps we mast take into consideration deformations of
torsion (\ref{torsdef}) and introduce $na_{(3)}$--parametrizations for $%
b(u,v)$ and ${P^\alpha }_{\beta \gamma }$ into the basic
na--equations (\ref {neq1}). The resulting equations, for
$na_{(3)}$--maps, are equivalent to equations (\ref{neqc3}) (with
a corresponding redefinition of deformation parameters). \qquad
$\Box $
\end{enumerate}

We point out that for ${\pi}_{(0)}$-maps we have not differential
equations on $P^{\alpha}_{{.}{\beta}{\gamma}}$ (in the isotropic
case one considers a first order system of differential equations
on metric \cite{sin}; we omit constructions with deformation of
metric in this Section).\

To formulate invariant conditions for reciprocal na--maps (when
every a-parallel on $\underline{{V}}{^{(n+m)}}$~ is also
transformed into na--parallel on $V^{(n+m)})$ it is convenient to
introduce into consideration
the curvature and Ricci tensors defined for auxiliary connection ${\gamma }_{%
{.}{\beta }{\gamma }}^\alpha $: $$
r_{{\alpha }{.}{\beta }{\tau }}^{{.}{\delta }}={\partial }_{[{\beta }}{%
\gamma }_{{.}{\tau }]{\alpha }}^\delta +{\gamma }_{{.}{\rho }[{\beta }%
}^\delta {\gamma }_{{.}{\tau }]{\alpha }}^\rho +{{\gamma }^\delta
}_{\alpha \phi }{w^\phi }_{\beta \tau } $$
and, respectively, $r_{{\alpha }{\tau }}=r_{{\alpha }{.}{\gamma }{\tau }}^{{.%
}{\gamma }}$, where $[\quad ]$ denotes antisymmetric alternation
of indices, and to define values:
\begin{eqnarray}
^{(0)}T_{{.}{\alpha }{\beta }}^\mu &= & {\Gamma }_{{.}{\alpha
}{\beta }}^\mu -T_{{.}{\alpha }{\beta }}^\mu
 -{\frac 1{(n+m+1)}}({\delta }_{({\alpha }}^\mu {%
\Gamma }_{{.}{\beta }){\delta }}^\delta -{\delta }_{({\alpha
}}^\mu T_{{.}{\beta }){\gamma }}^\delta ), \nonumber  \\ &{ } &
\nonumber \\ {}^{(0)}{W}_{\alpha \cdot \beta \gamma }^{\cdot \tau
} &=& {r}_{\alpha \cdot \beta \gamma }^{\cdot \tau }+{\frac
1{n+m+1}}[{\gamma }_{\cdot \varphi \tau
}^\tau {\delta }_{(\alpha }^\tau {w^\varphi }_{\beta )\gamma }-({\delta }%
_\alpha ^\tau {r}_{[\gamma \beta ]}+{\delta }_\gamma ^\tau
{r}_{[\alpha \beta ]}-{\delta }_\beta ^\tau {r}_{[\alpha \gamma
]})] \nonumber \\ &{ }& - {\frac 1{{(n+m+1)}^2}}
 [{\delta }_\alpha ^\tau (2{\gamma }_{\cdot \varphi
\tau }^\tau {w^\varphi }_{[\gamma \beta ]}
 -{\gamma }_{\cdot \tau [\gamma }^\tau {w^\varphi }_{\beta ]\varphi })
 \nonumber \\ &{}&
 +{\delta }_\gamma ^\tau (2{\gamma }%
_{\cdot \varphi \tau }^\tau {w^\varphi }_{\alpha \beta }-{\gamma
}_{\cdot
\alpha \tau }^\tau {w^\varphi }_{\beta \varphi })-{\delta }_\beta ^\tau (2{%
\gamma }_{\cdot \varphi \tau }^\tau {w^\varphi }_{\alpha \gamma }-{\gamma }%
_{\cdot \alpha \tau }^\tau {w^\varphi }_{\gamma \varphi })],
\nonumber \\
 &{ } & \nonumber \\
 {^{(3)}T}_{{.}{\alpha }{\beta }}^\delta & = &
 {\gamma }_{{.}{\alpha }{\beta }}^\delta +
{\frac 1{n+m}}({\delta }_\alpha ^\gamma -{\epsilon }{\varphi }%
^\delta q_\alpha )[{\gamma }_{{.}{\beta }{\tau }}^\tau +{\epsilon }{\varphi }%
^\tau {^{({\gamma })}D}_\beta q_\tau  \nonumber \\
& { }& + {\frac 1{{n+m-1}}}q_\beta ({\epsilon }{%
\varphi }^\tau {\gamma }_{{.}{\tau }{\lambda }}^\lambda +{\varphi }^\lambda {%
\varphi }^\tau {^{({\gamma })}D}_\tau q_\lambda )] +{\epsilon
}{\varphi }^\tau {^{({\gamma })}D}_\beta q_\tau  \nonumber \\
 &{ } & -{\frac 1{n+m}}({%
\delta }_\beta ^\delta - {\epsilon }{\varphi }^\delta q_\beta
)[{\gamma }_{{.}{\alpha }{\tau }}^\tau +{\epsilon }{\varphi }^\tau
{^{({\gamma })}D}_\alpha q_\tau  \nonumber \\
 & { } &
+{\frac 1{{n+m-1}}}q_\alpha ({\epsilon }{\varphi }^\tau {\gamma
}_{{.}{\tau }{\lambda }}^\lambda +
{\varphi }^\lambda {\varphi }^\tau {^{(\gamma )}D%
}_\tau q_\lambda )], \nonumber \\ &{ } & \nonumber \\
{^{(3)}W}^\alpha {{.}{\beta }{\gamma }{\delta }} & = &
{\rho }_{{\beta }{.}{\gamma } {\delta }}^{{.}{\alpha }}+{\epsilon }{\varphi }^\alpha %
 q_\tau {\rho }_{{\beta }{.}{\gamma }{\delta }}^{{.}{\tau }} \nonumber \\
& { } & +({\delta }_\delta ^\alpha -{%
\epsilon }{\varphi }^\alpha q_\delta )p_{{\beta }{\gamma }}-({\delta }%
_\gamma ^\alpha -{\epsilon }{\varphi }^\alpha q_\gamma )p_{{\beta }{\delta }}%
-({\delta }_\beta ^\alpha -{\epsilon }{\varphi }^\alpha q_\beta ) %
p_{[{\gamma }{\delta }]}, \nonumber  \\
 &{ }& \nonumber \\
(n+m-2)p_{{\alpha }{\beta }} &=&
 -{\rho }_{{\alpha }{\beta }}-{\epsilon }q_\tau {%
\varphi }^\gamma {\rho }_{{\alpha }{.}{\beta }{\gamma }}^{{.}{\tau
}}
 +{\frac
1{n+m}}[{\rho }_{{\tau }{.}{\beta }{\alpha }}^{{.}{\tau
}}-{\epsilon }q_\tau
{\varphi }^\gamma {\rho }_{{\gamma }{.}{\beta }{\alpha }}^{{.}{\tau }}+{%
\epsilon }q_\beta {\varphi }^\tau {\rho }_{{\alpha }{\tau }}
\nonumber \\
 &{ } &+
{\epsilon }q_\alpha (-{\varphi }^\gamma %
{\rho }_{{\tau }{.}{\beta }{\gamma }}^{{.}{\tau }}+%
{\epsilon }q_\tau {\varphi }^\gamma {\varphi }^\delta {\rho }%
_{{\gamma }{.}{\beta }{\delta }}^{{.}{\tau }}]), \nonumber
\end{eqnarray}
where $q_\alpha {\varphi }^\alpha ={\epsilon }=\pm 1,$ $$ {{\rho
}^\alpha }_{\beta \gamma \delta }=r_{\beta \cdot \gamma \delta
}^{\cdot \alpha }+{\frac 12}({\psi }_{(\beta }{\delta }_{\varphi )}^\alpha +{%
\sigma }_{\beta \varphi }{\varphi }^\tau ){w^\varphi }_{\gamma
\delta } $$ and ${\rho }_{\alpha \beta }={\rho }_{\cdot \alpha
\beta \tau }^\tau .$

For similar values on $\underline{{V}}{^{(n+m)}}$ we write, for
instance, $$
{\underline{\rho }}_{\cdot \beta \gamma \delta }^\alpha ={\underline{r}}%
_{\beta \cdot \gamma \delta }^{\cdot \alpha }-{\frac 12}({\psi }_{(\beta }{%
\delta }_{{\varphi })}^\alpha -{\sigma }_{\beta \varphi }{\varphi }^\tau ){%
w^\varphi }_{\gamma \delta }\ ) $$
and note that $^{(0)}{\underline{T}}_{{.}{\beta }{\gamma }}^\alpha ,^{(0)}{%
\underline{W}}_{{.}{\alpha }{\beta }{\gamma }}^\nu ,{\hat T}_{{.}{\beta }{%
\gamma }}^\alpha ,{\check T}_{{.}{\beta }{\tau }}^\alpha ,{\hat W}_{{.}{%
\alpha }{\beta }{\gamma }}^\delta ,{\check W}_{{.}{\alpha }{\beta }{\gamma }%
}^\delta ,^{(3)}{\underline{T}}_{{.}{\alpha }{\beta }}^\delta ,\ {}^{(3)}{%
\underline{W}}_{{.}{\beta }{\gamma }{\delta }}^\alpha $ are given,
correspondingly, by auxiliary connections ${\underline{\Gamma
}}_{{.}{\alpha
}{\beta }}^\mu ,$%
$$
{\star {\gamma }}_{{.}{\beta }{\lambda }}^\alpha ={\gamma }_{{.}{\beta }{%
\lambda }}^\alpha +{\epsilon }F_\tau ^\alpha {^{({\gamma })}D}_{({\beta }}F_{%
{\lambda })}^\tau ,\quad {\check \gamma }_{{.}{\beta }{\lambda }}^\alpha ={%
\widetilde{\gamma }}_{{.}{\beta }{\lambda }}^\alpha +{\epsilon
}F_\tau ^\lambda {\widetilde{D}}_{({\beta }}F_{{\lambda })}^\tau ,
$$ $$
{\widetilde{\gamma }}_{{.}{\beta }{\tau }}^\alpha ={\gamma }_{{.}{\beta }{%
\tau }}^\alpha +{\sigma }_{({\beta }}F_{{\tau })}^\alpha ,\quad
{\hat \gamma
}_{{.}{\beta }{\lambda }}^\alpha ={\star {\gamma }}_{{.}{\beta }{\lambda }%
}^\alpha +{\widetilde{\sigma }}_{({\beta }}{\delta }_{{\lambda
})}^\alpha , $$ where ${\widetilde{\sigma }}_\beta ={\sigma
}_\alpha F_\beta ^\alpha .$

\begin{theorem}
Four classes of reciprocal na--maps of la--spacetimes are
characterized by corresponding invariant criterions:

\begin{enumerate}
\item  for a--maps
\begin{eqnarray}
^{(0)}T_{{.}{\alpha }{\beta }}^\mu &=&
^{(0)}{\underline{T}}_{{.}{\alpha }{\beta }}^\mu , \label{nainv0}
\\ {}^{(0)}W_{{.}{\alpha }{\beta }{\gamma }}^\delta & =&
^{(0)}{\underline{W}}_{{.}{\alpha }{\beta }{\gamma }}^\delta ;
 \nonumber
\end{eqnarray}

\item  for $na_{(1)}$--maps
\begin{eqnarray}
3({^{({\gamma })}D}_\lambda P_{{.}{\alpha }{\beta }}^\delta &+ &
P_{{.}{\tau }{\lambda }}^\delta P_{{.}{\alpha }{\beta }}^\tau )=
r_{({\alpha }{.}{\beta }){\lambda }}^{{.}{\delta }}-
{\underline{r}}_{({\alpha }{.}{\beta }){\lambda }}^{{.}{\delta }}
\nonumber \\ &{}& + [T_{{.}{\tau }({\alpha }}^\delta P_{{.}{\beta
}{\lambda })}^\tau + Q_{{.}{\tau }({\alpha }}^\delta P_{{.}{\beta
}{\lambda })}^\tau + b_{({\alpha }}P_{{.}{\beta }{\lambda
})}^\delta +{\delta }_{({\alpha }}^\delta a_{{\beta }{\lambda
})}]; \label{nainv1}
\end{eqnarray}

\item  for $na_{(2)}$--maps
\begin{eqnarray}
{\hat T}_{{.}{\beta }{\tau }}^\alpha &= & {\star T}_{{.}{\beta
}{\tau }}^\alpha , \label{nainv2} \\ {\hat W}_{{.}{\alpha }{\beta
}{\gamma }}^\delta & = & {\star W}_{{.}{\alpha }{\beta }{\gamma
}}^\delta ; \nonumber
\end{eqnarray}

\item  for $na_{(3)}$--maps
\begin{eqnarray}
{}^{(3)}T_{{.}{\beta }{\gamma }}^\alpha &=&
 {}^{(3)}{\underline{T}}_{{.}{\beta }{\gamma }}^\alpha , \label{nainv3} \\
{}^{(3)}W_{{.}{\beta }{\gamma }{\delta }}^\alpha &=&
{}^{(3)}{\underline{W}}_{{.}{\beta }{\gamma }{\delta }}^\alpha
.\nonumber
\end{eqnarray}
\end{enumerate}
\end{theorem}

{\sf Proof. }

\begin{enumerate}
\item  Let us prove that the a--invariant conditions (\ref{nainv0}) hold.
Deformations of d--connections of type
\begin{equation}
\label{auxc1}{}^{(0)}{\underline{\gamma }}_{\cdot \alpha \beta }^\mu ={{%
\gamma }^\mu }_{\alpha \beta }+{\psi }_{(\alpha }{\delta }_{\beta
)}^\mu
\end{equation}
define a--applications. Contracting indices $\mu $ and $\beta $ we
can write
\begin{equation}
\label{auxc2}{\psi }_\alpha ={\frac 1{m+n+1}}({{\underline{\gamma }}^\beta }%
_{\alpha \beta }-{{\gamma }^\beta }_{\alpha \beta }).
\end{equation}
Introducing the d--vector ${\psi }_\alpha $ into previous relation
and expressing $$
{{\gamma }^\alpha }_{\beta \tau }=-{T^\alpha }_{\beta \tau }+{{\Gamma }%
^\alpha }_{\beta \tau } $$ and similarly for underlined values we
obtain the first invariant conditions from (\ref{nainv0}).

Putting deformation (\ref{auxc1}) into the formulas for ${\underline{r}}%
_{\alpha \cdot \beta \gamma }^{\cdot \tau }$ and
${\underline{r}}_{\alpha \beta }={\underline{r}}_{\alpha \tau
\beta \tau }^{\cdot \tau }$ we obtain respectively the relations
\begin{equation}
\label{auxc3}{\underline{r}}_{\alpha \cdot \beta \gamma }^{\cdot
\tau
}-r_{\alpha \cdot \beta \gamma }^{\cdot \tau }={\delta }_\alpha ^\tau {\psi }%
_{[\gamma \beta ]}+{\psi }_{\alpha [\beta }{\delta }_{\gamma ]}^\tau +{%
\delta }_{(\alpha }^\tau {\psi }_{\varphi )}{w^\varphi }_{\beta
\gamma }
\end{equation}
and
\begin{equation}
\label{auxc4}{\underline{r}}_{\alpha \beta }-r_{\alpha \beta }={\psi }%
_{[\alpha \beta ]}+(n+m-1){\psi }_{\alpha \beta }+{\psi }_\varphi
{w^\varphi }_{\beta \alpha }+{\psi }_\alpha {w^\varphi }_{\beta
\varphi },
\end{equation}
where $$
{\psi }_{\alpha \beta }={}^{({\gamma })}D_\beta {\psi }_\alpha -{\psi }%
_\alpha {\psi }_\beta . $$
Putting (\ref{auxc1}) into (\ref{auxc3}) and (\ref{auxc4}) we can express ${%
\psi }_{[\alpha \beta ]}$ as
\begin{eqnarray}
&{}&{\psi }_{[\alpha \beta ]} =
{\frac 1{n+m+1}}[{\underline{r}}_{[\alpha \beta ]}+%
{\frac 2{n+m+1}}{\underline{\gamma }}_{\cdot \varphi \tau }^\tau {w^\varphi }%
_{[\alpha \beta ]}-{\frac 1{n+m+1}}{\underline{\gamma }}_{\cdot
\tau [\alpha }^\tau {w^\varphi }_{\beta ]\varphi }]  \nonumber \\
 & - & {\frac 1{n+m+1}}[r_{[\alpha \beta ]}+
{\frac 2{n+m+1}}{{\gamma }^\tau }_{\varphi \tau }
{w^\varphi }_{[\alpha \beta ]}-{\frac 1{n+m+1}}{{\gamma }%
^\tau }_{\tau [\alpha }{w^\varphi }_{\beta ]\varphi }].
\label{auxc5}
\end{eqnarray}
To simplify our consideration we can choose an a--transform,
pa\-ra\-met\-riz\-ed by corresponding $\psi $--vector from
(\ref{auxc1}), (or fix a local coordinate cart) the
antisymmetrized relations (\ref{auxc1}) to be satisfied by
d--tensor
\begin{eqnarray}
{\psi }_{\alpha \beta } &=&
{\frac 1{n+m+1}}[{\underline{r}}_{\alpha \beta }+{%
\frac 2{n+m+1}}{\underline{\gamma }}_{\cdot \varphi \tau }^\tau {w^\varphi }%
_{\alpha \beta }-{\frac 1{n+m+1}}{\underline{\gamma }}_{\cdot
\alpha \tau }^\tau {w^\varphi }_{\beta \varphi } \nonumber \\ &{}&
-
r_{\alpha \beta }-{\frac 2{n+m+1}}{{\gamma }^\tau }_{\varphi \tau }{%
w^\varphi }_{\alpha \beta }+{\frac 1{n+m+1}}{{\gamma }^\tau }_{\alpha \tau }{%
w^\varphi }_{\beta \varphi }]\label{auxc6}
\end{eqnarray}
Introducing expressions (\ref{auxc1}), (\ref{auxc5}) and
(\ref{auxc6}) into
deformation of curvature (\ref{auxc2}) we obtain the second condition from (%
\ref{nainv0}) of a-map invariance: $$
^{(0)}W_{\alpha \cdot \beta \gamma }^{\cdot \delta }={}^{(0)}{\underline{W}}%
_{\alpha \cdot \beta \gamma }^{\cdot \delta }, $$ where the Weyl
d--tensor on $\underline{{V}}{^{(n+m)}}$ is defined as
\begin{eqnarray}
{}^{(0)}{\underline{W}}_{\alpha \cdot \beta \gamma }^{\cdot \tau }& = &{%
\underline{r}}_{\alpha \cdot \beta \gamma }^{\cdot \tau }+{\frac 1{n+m+1}}[{%
\underline{\gamma }}_{\cdot \varphi \tau }^\tau {\delta }_{(\alpha }^\tau {%
w^\varphi }_{\beta )\gamma }-({\delta }_\alpha ^\tau {\underline{r}}%
_{[\gamma \beta ]}+{\delta }_\gamma ^\tau {\underline{r}}_{[\alpha \beta ]}-{%
\delta }_\beta ^\tau {\underline{r}}_{[\alpha \gamma ]})]
\nonumber \\
 &{ } &
-{\frac 1{{(n+m+1)}^2}}[{\delta }_\alpha ^\tau (2{\underline{\gamma }}%
_{\cdot \varphi \tau }^\tau {w^\varphi }_{[\gamma \beta ]}-
{\underline{\gamma }}_{\cdot \tau [\gamma }^\tau {w^\varphi
}_{\beta ]\varphi })
 \nonumber \\
 & { } &
+{\delta }_\gamma ^\tau (2{\underline{\gamma }}_{\cdot \varphi \tau }^\tau {%
w^\varphi }_{\alpha \beta }-{\underline{\gamma }}_{\cdot \alpha \tau }^\tau {%
w^\varphi }_{\beta \varphi })-{\delta }_\beta ^\tau (2{\underline{\gamma }}%
_{\cdot \varphi \tau }^\tau {w^\varphi }_{\alpha \gamma
}-{\underline{\gamma }}_{\cdot \alpha \tau }^\tau {w^\varphi
}_{\gamma \varphi })].
 \nonumber
\end{eqnarray}

\item  To obtain $na_{(1)}$--invariant conditions we rewrite $na_{(1)}$%
--equations (\ref{neqc1}) as to consider in explicit form
covariant derivation $^{({\gamma })}D$ and deformations
(\ref{simdef}) and (\ref {torsdef}):
\begin{eqnarray}
&2&({}^{({\gamma })}D_\alpha {P^\delta }_{\beta \gamma }+
{}^{({\gamma })}D_\beta {P^\delta }_{\alpha \gamma }+{}^{({\gamma })}D_\gamma {P^\delta }%
_{\alpha \beta }+{P^\delta }_{\tau \alpha }{P^\tau }_{\beta \gamma }+{%
P^\delta }_{\tau \beta }{P^\tau }_{\alpha \gamma } \nonumber \\
 &{}& +{P^\delta }_{\tau \gamma }{P^\tau }_{\alpha \beta })  =
{T^\delta }_{\tau (\alpha }{P^\tau }_{\beta \gamma )}+{H^\delta
}_{\tau (\alpha }{P^\tau }_{\beta \gamma )}+b_{(\alpha }{P^\delta
}_{\beta \gamma )}+a_{(\alpha \beta }{\delta }_{\gamma )}^\delta
.\label{nainv1c}
\end{eqnarray}
Alternating the first two indices in (\ref{nainv1c}) we have
\begin{eqnarray}
2({\underline{r}}_{(\alpha \cdot \beta )\gamma }^{\cdot \delta
}-r_{(\alpha \cdot \beta )\gamma }^{\cdot \delta }) &=&
2({}^{(\gamma )}D_\alpha {P^\delta }%
_{\beta \gamma }+{}^{(\gamma )}D_\beta {P^\delta }_{\alpha \gamma
}-2{}^{(\gamma )}D_\gamma {P^\delta }_{\alpha \beta }  \nonumber
\\ &{}& +{P^\delta }_{\tau \alpha }{P^\tau }_{\beta \gamma
}+{P^\delta }_{\tau \beta }{P^\tau }_{\alpha \gamma }-2{P^\delta
}_{\tau \gamma }{P^\tau }_{\alpha \beta }). \nonumber
\end{eqnarray}
Substituting the last expression from (\ref{nainv1c}) and
rescalling the deformation parameters and d--tensors we obtain the
conditions (\ref{neqc1}).

\item  Now we prove the invariant conditions for $na_{(0)}$--maps satisfying
conditions $$ \epsilon \neq 0\quad \mbox{and}\quad \epsilon
-F_\beta ^\alpha F_\alpha ^\beta \neq 0 $$ Let define the
auxiliary d--connection
\begin{equation}
\label{auxc7}{\tilde \gamma }_{\cdot \beta \tau }^\alpha
={\underline{\gamma
}}_{\cdot \beta \tau }^\alpha -{\psi }_{(\beta }{\delta }_{\tau )}^\alpha ={{%
\gamma }^\alpha }_{\beta \tau }+{\sigma }_{(\beta }F_{\tau
)}^\alpha
\end{equation}
and write $$
{\tilde D}_\gamma ={}^{({\gamma })}D_\gamma F_\beta ^\alpha +{\tilde \sigma }%
_\gamma F_\beta ^\alpha -{\epsilon }{\sigma }_\beta {\delta
}_\gamma ^\alpha , $$ where ${\tilde \sigma }_\beta ={\sigma
}_\alpha F_\beta ^\alpha ,$ or, as a consequence from the last
equality, $$
{\sigma }_{(\alpha }F_{\beta )}^\tau ={\epsilon }F_\lambda ^\tau ({}^{({%
\gamma })}D_{(\alpha }F_{\beta )}^\alpha -{\tilde D}_{(\alpha
}F_{\beta )}^\lambda )+{\tilde \sigma }_{(}{\alpha }{\delta
}_{\beta )}^\tau . $$ Introducing the auxiliary connections $$
{\star {\gamma }}_{\cdot \beta \lambda }^\alpha ={\gamma }_{\cdot
\beta \lambda }^\alpha +{\epsilon }F_\tau ^\alpha {}^{({\gamma
})}D_{(\beta }F_{\lambda )}^\tau ,{\check \gamma }_{\cdot \beta
\lambda }^\alpha ={\tilde
\gamma }_{\cdot \beta \lambda }^\alpha +{\epsilon }F_\tau ^\alpha {\tilde D}%
_{(\beta }F_{\lambda )}^\tau $$ we can express the deformation
(\ref{auxc7}) in a form characteristic for a--maps:
\begin{equation}
\label{auxc8}{\hat \gamma }_{\cdot \beta \gamma }^\alpha ={\star {\gamma }}%
_{\cdot \beta \gamma }^\alpha +{\tilde \sigma }_{(\beta }{\delta
}_{\lambda )}^\alpha .
\end{equation}
Now it's obvious that $na_{(2)}$--invariant conditions
(\ref{auxc8}) are equivalent with a--invariant conditions
(\ref{nainv0}) written for d--connection (\ref{auxc8}). As a
matter of principle we can write formulas for such
$na_{(2)}$--invariants in terms of ''underlined'' and
''non--underlined'' values by expressing consequently all used
auxiliary connections as deformations of ''prime'' connections on
${V^{(n+m)}}$ and ''final'' connections on
$\underline{{V}}{^{(n+m)}}.$ We omit such tedious calculations in
this work.

\item  Finally, we prove the last statement, for $na_{(3)}$--maps, of this
theorem. Let
\begin{equation}
\label{auxc9}q_\alpha {\varphi }^\alpha =e=\pm 1,
\end{equation}
where ${\varphi }^\alpha $ is contained in
\begin{equation}
\label{auxc10}{\underline{\gamma }}_{\cdot \beta \gamma }^\alpha ={{\gamma }%
^\alpha }_{\beta \gamma }+{\psi }_{(\beta }{\delta }_{\gamma )}^\alpha +{%
\sigma }_{\beta \gamma }{\varphi }^\alpha .
\end{equation}
Acting with operator $^{({\gamma })}{\underline{D}}_\beta $ on
(\ref{auxc8}) we write
\begin{equation}
\label{auxc11}{}^{({\gamma })}{\underline{D}}_\beta q_\alpha ={}^{({\gamma }%
)}D_\beta q_\alpha -{\psi }_{(\alpha }q_{\beta )}-e{\sigma
}_{\alpha \beta }.
\end{equation}
Contracting (\ref{auxc10}) with ${\varphi }^\alpha $ we can
express $$ e{\varphi }^\alpha {\sigma }_{\alpha \beta }={\varphi
}^\alpha ({}^{({\gamma
})}D_\beta q_\alpha -{}^{({\gamma })}{\underline{D}}_\beta q_\alpha )-{%
\varphi }_\alpha q^\alpha q_\beta -e{\psi }_\beta . $$ Putting the
last formula in (\ref{auxc9}) contracted on indices $\alpha $ and
$\gamma $ we obtain
\begin{equation}
\label{auxc12}(n+m){\psi }_\beta ={\underline{\gamma }}_{\cdot
\alpha \beta
}^\alpha -{{\gamma }^\alpha }_{\alpha \beta }+e{\psi }_\alpha {\varphi }%
^\alpha q_\beta +e{\varphi }^\alpha {\varphi }^\beta ({}^{({\gamma })}{%
\underline{D}}_\beta -{}^{({\gamma })}D_\beta ).
\end{equation}
>From these relations, taking into consideration (\ref{auxc8}), we have%
$$
(n+m-1){\psi }_\alpha {\varphi }^\alpha ={\varphi }^\alpha ({\underline{%
\gamma }}_{\cdot \alpha \beta }^\alpha -{{\gamma }^\alpha
}_{\alpha \beta
})+e{\varphi }^\alpha {\varphi }^\beta ({}^{({\gamma })}{\underline{D}}%
_\beta q_\alpha -{}^{({\gamma })}D_\beta q_\alpha ) $$

Using the equalities and identities (\ref{auxc10}) and
(\ref{auxc11}) we can express the deformations (\ref{auxc9}) as
the first $na_{(3)}$--invariant conditions from (\ref{nainv3}).

To prove the second class of $na_{(3)}$--invariant conditions we
introduce two additional d--tensors: $$ {{\rho }^\alpha }_{\beta
\gamma \delta }=r_{\beta \cdot \gamma \delta
}^{\cdot \alpha }+{\frac 12}({\psi }_{(\beta }{\delta }_{\varphi )}^\alpha +{%
\sigma }_{\beta \varphi }{\varphi }^\tau ){w^\varphi }_{\gamma
\delta } $$ and
\begin{equation}
\label{rho1}{\underline{\rho }}_{\cdot \beta \gamma \delta }^\alpha ={%
\underline{r}}_{\beta \cdot \gamma \delta }^{\cdot \alpha }-{\frac
12}({\psi
}_{(\beta }{\delta }_{{\varphi })}^\alpha -{\sigma }_{\beta \varphi }{%
\varphi }^\tau ){w^\varphi }_{\gamma \delta }.
\end{equation}
Using deformation (\ref{auxc9}) and (\ref{auxc12}) we write
relation
\begin{equation}
\label{sigma1}{\tilde \sigma }_{\cdot \beta \gamma \delta }^\alpha ={%
\underline{\rho }}_{\cdot \beta \gamma \delta }^\alpha -{\rho
}_{\cdot \beta
\gamma \delta }^\alpha ={\psi }_{\beta [\delta }{\delta }_{\gamma ]}^\alpha -%
{\psi }_{[{\gamma }{\delta }]}{\delta }_\beta ^\alpha -{\sigma
}_{\beta \gamma \delta }{\varphi }^\alpha ,
\end{equation}
where $$
{\psi }_{\alpha \beta }={}^{({\gamma })}D_\beta {\psi }_\alpha +{\psi }%
_\alpha {\psi }_\beta -({\nu }+{\varphi }^\tau {\psi }_\tau ){\sigma }%
_{\alpha \beta }, $$ and $$
{\sigma }_{\alpha \beta \gamma }={}^{({\gamma })}D_{[\gamma }{\sigma }_{{%
\beta }]{\alpha }}+{\mu }_{[\gamma }{\sigma }_{{\beta }]{\alpha }}-{\sigma }%
_{{\alpha }[{\gamma }}{\sigma }_{{\beta }]{\tau }}{\varphi }^\tau
. $$ Let multiply (\ref{rho1}) on $q_\alpha $ and write (taking
into account relations (\ref{auxc8})) the relation
\begin{equation}
\label{sigma2}e{\sigma }_{\alpha \beta \gamma }=-q_\tau {\tilde \sigma }%
_{\cdot \alpha \beta \delta }^\tau +{\psi }_{\alpha [\beta }q_{\gamma ]}-{%
\psi }_{[\beta \gamma ]}q_\alpha .
\end{equation}
The next step is to express ${\psi }_{\alpha \beta }$ trough d--objects on ${%
V^{(n+m)}}.$ To do this we contract indices $\alpha $ and $\beta $
in (\ref {rho1}) and obtain $$ (n+m){\psi }_{[\alpha \beta
]}=-{\sigma }_{\cdot \tau \alpha \beta }^\tau
+eq_\tau {\varphi }^\lambda {\sigma }_{\cdot \lambda \alpha \beta }^\tau -e{%
\tilde \psi }_{[\alpha }{\tilde \psi }_{\beta ]}. $$
Then contracting indices $\alpha $ and $\delta $ in (\ref{rho1}) and using (%
\ref{sigma1}) we write
\begin{equation}
\label{auxc13}(n+m-2){\psi }_{\alpha \beta }={\tilde \sigma
}_{\cdot \alpha \beta \tau }^\tau -eq_\tau {\varphi }^\lambda
{\tilde \sigma }_{\cdot \alpha \beta \lambda }^\tau +{\psi
}_{[\beta \alpha ]}+e({\tilde \psi }_\beta q_\alpha -{\hat \psi
}_{(\alpha }q_{\beta )}),
\end{equation}
where ${\hat \psi }_\alpha ={\varphi }^\tau {\psi }_{\alpha \tau
}.$ If the both parts of (\ref{sigma2}) are contracted with
${\varphi }^\alpha ,$ it results that $$ (n+m-2){\tilde \psi
}_\alpha ={\varphi }^\tau {\sigma }_{\cdot \tau \alpha
\lambda }^\lambda -eq_\tau {\varphi }^\lambda {\varphi }^\delta {\sigma }%
_{\lambda \alpha \delta }^\tau -eq_\alpha , $$ and, in consequence
of ${\sigma }_{\beta (\gamma \delta )}^\alpha =0,$ we have $$
(n+m-1){\varphi }={\varphi }^\beta {\varphi }^\gamma {\sigma
}_{\cdot \beta \gamma \alpha }^\alpha . $$ By using the last
expressions we can write
\begin{equation}
\label{auxc14}(n+m-2){\underline{\psi }}_\alpha ={\varphi }^\tau {\sigma }%
_{\cdot \tau \alpha \lambda }^\lambda -eq_\tau {\varphi }^\lambda {\varphi }%
^\delta {\sigma }_{\cdot \lambda \alpha \delta }^\tau -e{(n+m-1)}%
^{-1}q_\alpha {\varphi }^\tau {\varphi }^\lambda {\sigma }_{\cdot
\tau \lambda \delta }^\delta .
\end{equation}
Contracting (\ref{sigma2}) with ${\varphi }^\beta $ we have $$
(n+m){\hat \psi }_\alpha ={\varphi }^\tau {\sigma }_{\cdot \alpha
\tau \lambda }^\lambda +{\tilde \psi }_\alpha $$
and taking into consideration (\ref{auxc13}) we can express ${\hat \psi }%
_\alpha $ through ${\sigma }_{\cdot \beta \gamma \delta }^\alpha
.$

As a consequence of (\ref{sigma1})--(\ref{auxc13}) we obtain this
formulas for d--tensor ${\psi }_{\alpha \beta }:$
\begin{eqnarray} &{}&
(n+m-2){\psi }_{\alpha \beta }={\sigma }_{\cdot \alpha \beta \tau
}^\tau -eq_\tau {\varphi }^\lambda {\sigma }_{\cdot \alpha \beta
\lambda }^\tau
 \nonumber \\ &{}&
+ {\frac 1{n+m}}\{-{\sigma }_{\cdot \tau \beta \alpha }^\tau +
eq_\tau {\varphi }^\lambda {\sigma }_{\cdot \lambda \beta \alpha
}^\tau - q_\beta (e{\varphi }^\tau {\sigma }_{\cdot \alpha \tau
\lambda }^\lambda - q_\tau {\varphi }^\lambda {\varphi }^\delta
{\sigma }_{\cdot \alpha \lambda \delta }^\tau )+eq_\alpha \times
\nonumber  \\
 &{}&
[{\varphi }^\lambda {\sigma }_{\cdot \tau \beta \lambda }^\tau -eq_\tau {%
\varphi }^\lambda {\varphi }^\delta {\sigma }_{\cdot \lambda \beta
\delta
}^\tau -{\frac e{n+m-1}}q_\beta ({\varphi }^\tau {\varphi }^\lambda {\sigma }%
_{\cdot \tau \gamma \delta }^\delta -eq_\tau {\varphi }^\lambda {\varphi }%
^\delta {\varphi }^\varepsilon {\sigma }_{\cdot \lambda \delta
\varepsilon }^\tau )]\}. \nonumber
\end{eqnarray}

Finally, putting the last formula and (\ref{sigma1}) into
(\ref{rho1}) and
after a rearrangement of terms we obtain the second group of $na_{(3)}$%
-invariant conditions (\ref{nainv3}). If necessary we can rewrite
these
conditions in terms of geometrical objects on ${V^{(n+m)}}$ and $\underline{{V}%
}{^{(n+m)}}.$ To do this we mast introduce splittings
(\ref{auxc12}) into (\ref {nainv3}). \qquad $\Box $
\end{enumerate}

For the particular case of $na_{(3)}$--maps when ${\psi }_\alpha
=0,{\varphi
}_\alpha =g_{\alpha \beta }{\varphi }^\beta ={\frac \delta {\delta u^\alpha }%
}(\ln {\Omega }),$ ${\Omega }(u)>0$ and ${\sigma }_{\alpha \beta
}=g_{\alpha
\beta }$ we define a subclass of conformal transforms ${\underline{g}}%
_{\alpha \beta }(u)={\Omega }^2(u)g_{\alpha \beta }$ which, in
consequence of the fact that d--vector ${\varphi }_\alpha $ must
satisfy equations (\ref {neqc3}), generalizes the class of
concircular transforms (see \cite{sin} for references and details
on concircular mappings of Riemannaian spaces).

We emphasize that the basic na--equations
(\ref{neqc1})--(\ref{neqc3}) are systems of first order partial
differential equations. The study of their geometrical properties
and definition of integral varieties, general and
particular solutions are possible by using the formalism of Pffaf systems%
\cite{car45,voa,vm}. Here we point out that by using algebraic
methods we can always verify if systems of na--equations of type
(\ref{neqc1})--(\ref {neqc3}) are, or not, involute, even to find
their explicit solutions it is a difficult task (see more detailed
considerations for isotropic ng--maps in
\cite{sin} and, on language of Pffaf systems for na--maps, in \cite{vm,voa}%
). We can also formulate the Cauchy problem for na--equations on ${V^{(n+m)}}$%
~ and choose deformation parameters (\ref{defpar}) as to make
involute
mentioned equations for the case of maps to a given background space $%
\underline{{V}}{^{(n+m)}}$. If a solution, for example, of
$na_{(1)}$--map
equations exists, we say that the la--spacetime ${V^{(n+m)}}$ is $na_{(1)}$%
--projective to the la--spacetime $\underline{{V}}{^{(n+m)}}.$ In
general, we have to introduce chains of na--maps in order to
obtain involute systems of
equations for maps (superpositions of na-maps) from ${V^{(n+m)}}$ to $%
\underline{{V}}{^{(n+m)}}:$
$$U\ {\buildrel {ng<i_{1}>} \over \longrightarrow}\  {U_{\underline 1}}\ %
{\buildrel ng<i_2> \over \longrightarrow}\ \cdots \ %
{\buildrel ng<i_{k-1}> \over \longrightarrow}\ U_{\underline {k-1}} \ %
{\buildrel ng<i_k> \over \longrightarrow}\ {\underline U} $$ %
%
where $$ U\subset {V^{(n+m)}},U_{\underline{1}}\subset
{V^{(n+m)}}_{\underline{1}},\ldots
,U_{\underline{k-1}}\subset {V^{(n+m)}}_{\underline{k-1}},U_{\underline{k}%
}\subset {V^{(n+m)}}_{\underline{k}},{\underline{U}}\subset
{V^{(n+m)}} $$ with corresponding splittings of auxiliary
symmetric connections $$
{\underline{\gamma }}_{.{\beta }{\gamma }}^\alpha =_{<i_1>}P_{.{\beta }{%
\gamma }}^\alpha +_{<i_2>}P_{.{\beta }{\gamma }}^\alpha +\cdots +_{<i_k>}P_{.%
{\beta }{\gamma }}^\alpha $$ and torsion $$
{\underline{T}}_{.{\beta }{\gamma }}^\alpha =T_{.{\beta }{\gamma
}}^\alpha
+_{<i_1>}Q_{.{\beta }{\gamma }}^\alpha +_{<i_2>}Q_{.{\beta }{\gamma }%
}^\alpha +\cdots +_{<i_k>}Q_{.{\beta }{\gamma }}^\alpha $$ where
the indices $<i_1>=0,1,2,3,$ denote possible types of na--maps.

\begin{definition}
A la--spacetime ${V^{(n+m)}}$~ is nearly conformally projective to
the
la--spa\-ce\-ti\-me $\underline{{V}}{^{(n+m)}}, nc:{V^{(n+m)}\to }\underline{{V}}{%
^{(n+m)}},$~ if there is a finite chain of na--maps from ${V^{(n+m)}}$~ to $%
\underline{{V}}{^{(n+m)}}.$
\end{definition}

For nearly conformal maps we formulate:

\begin{theorem}
For every fixed triples $(N_j^a,{\Gamma }_{{.}{\beta }{\gamma
}}^\alpha
,U\subset {V^{(n+m)}})$ and $(N_j^a,{\underline{\Gamma }}_{{.}{\beta }{\gamma }%
}^\alpha $, ${\underline{U}}\subset \underline{{V}}{^{(n+m)}})$
and given components of nonlinear connection, d--connection and
d--metric being of class $C^r(U),$ $C^r({\underline{U}})$, $r>3,$
there is a finite chain of na--maps $nc: U\to {\underline{U}}.$
\end{theorem}

The proof is to performed by introducing a finite number of
na-maps with corresponding components of deformation parameters
and deformation tensors in order to transform step by step the
coefficients of d-connection ${\Gamma }_{\gamma \delta }^\alpha $
into the ${\underline{\Gamma }}_{\beta \gamma }^\alpha ).$

Now we introduce the concept of the Category of la--spacetimes, ${\cal C}({%
V^{(n+m)}}).$ The elements of ${\cal C}({V^{(n+m)}})$ consist from
objects $$Ob{\cal
C}({V^{(n+m)}})=\{{V^{(n+m)}},{V^{(n+m)}}_{<i_1>},{V^{(n+m)}}_{<i_2>},{\ldots
}\}$$
 being la--spacetimess, for simplicity in this work, having common
N--connection structures, and morphisms $Mor{\cal C}({V^{(n+m)}})=\{nc({V^{(n+m)}%
}_{<i_1>},{V^{(n+m)}}_{<i_2>})\}$ being chains of na--maps
interrelating la--spacetimes. We point out that we can consider
equivalent models of physical theories on every object of ${\cal
C}({V^{(n+m)}}).$ One of the main purposes of this section is to
develop a d--tensor and d--variational formalism on ${\cal
C}({V^{(n+m)}}),$ i.e. on la--multispaces, interrelated with
nc--maps. Taking into account the distinguished character of
geometrical objects on la--spacetimes we call tensors on ${\cal C}({V^{(n+m)}}%
) $ as distinguished tensors on la--spacetime Category, or
dc--tensors.

Finally, we emphasize that the presented in this Section
definitions and theorems can be generalized for (super) vector
bundles with arbitrary given structures of nonlinear connection,
linear d--connection and metric structures \cite{vm,voa}.

\section{The Nearly Autoparallel Tensor-Integral}

The aim of this Section is to define the tensor integration not
only for bitensors, objects defined on the same curved space, but
for dc--tensors, defined on two spaces, ${V^{(n+m)}}$ and
$\underline{{V}}{{^{(n+m)}}}$, even it is necessary on
la--multispaces. A. Mo\'or tensor--integral formalism \cite {moo}
having a lot of applications in classical and quantum gravity
\cite {syn,dewd,goz} was extended for locally isotropic
multispaces in \cite {vog,vob12}. The unispacial locally
anisotropic version is given in \cite {vm,vrjp,v295a,gv}.

Let $T_u{V^{(n+m)}}$~ and
$T_{\underline{u}}\underline{{V}}{{^{(n+m)}}}$ be
tangent spaces in corresponding points $u{\in }$ $U{\subset V^{(n+m)}}$ and ${%
\underline{u}}{\in }{\underline{U}}{\subset
}\underline{{V}}{{^{(n+m)}}}$ and,
respectively, $T_u^{*}{V^{(n+m)}}$ and $T_{\underline{u}}^{*}\underline{{V}}{{%
^{(n+m)}}}$ be their duals (in general, in this Section we shall
not consider
that a common coordinatization is introduced for open regions $U$ and ${%
\underline{U}}$ ). We call as the dc--tensors on the pair of spaces $({%
V^{(n+m)}},\underline{{V}}{{^{(n+m)}}})$ the elements of
distinguished tensor algebra $$
({\otimes }_\alpha T_u{V^{(n+m)}}){\otimes }({\otimes }_\beta T_u^{*}{V^{(n+m)}})%
{\otimes }({\otimes }_\gamma T_{\underline{u}}\underline{{V}}{{^{(n+m)}}}){%
\otimes }({\otimes }_\delta
T_{\underline{u}}^{*}\underline{{V}}{{^{(n+m)}}}) $$ defined over
the space ${V^{(n+m)}}\otimes \underline{{V}}{{^{(n+m)}}},$ for a
given $nc:{V^{(n+m)}}\rightarrow \underline{{V}}{{^{(n+m)}}}.$

We admit the convention that underlined and non--underlined
indices refer,
respectively, to the points ${\underline{u}}$ and $u$. Thus $Q_{{.}{%
\underline{\alpha }}}^\beta ,$ for instance, are the components of
dc--tensor $Q{\in }T_u{V^{(n+m)}\otimes }T_{\underline{u}}\underline{{V}}{{%
^{(n+m)}}}.$

Now, we define the transport dc--tensors. Let open regions $U$ and ${%
\underline{U}}$ be homeomorphic to sphere ${\cal R}^{2(n+m)}$ and
introduce
isomorphism ${\mu }_{{u},{\underline{u}}}$ between $T_u{V^{(n+m)}}$ and $T_{%
\underline{u}}\underline{{V}}{{^{(n+m)}}}$ (given by map $nc:U{\to }{%
\underline{U}}).$ We consider that for every d--vector $v^\alpha {\in }T_u{%
V^{(n+m)}}$ corresponds the vector ${\mu }_{{u},{\underline{u}}}(v^\alpha )=v^{%
\underline{\alpha }}{\in
}T_{\underline{u}}\underline{{V}}{{^{(n+m)}}},$ with components
$v^{\underline{\alpha }}$ being linear functions of $v^\alpha $:
$$
v^{\underline{\alpha }}=h_\alpha ^{\underline{\alpha }}(u,{\underline{u}}%
)v^\alpha ,\quad v_{\underline{\alpha }}=h_{\underline{\alpha }}^\alpha ({%
\underline{u}},u)v_\alpha , $$ where $h_{\underline{\alpha
}}^\alpha ({\underline{u}},u)$ are the components of dc--tensor
associated with ${\mu }_{u,{\underline{u}}}^{-1}$. In a similar
manner we have $$
v^\alpha =h_{\underline{\alpha }}^\alpha ({\underline{u}},u)v^{\underline{%
\alpha }},\quad v_\alpha =h_\alpha ^{\underline{\alpha }}(u,{\underline{u}}%
)v_{\underline{\alpha }}. $$

In order to reconcile just presented definitions and to assure the
identity for trivial maps ${V^{(n+m)}\to
}\underline{{V}}{{^{(n+m)}}},u={\underline{u}},$ the transport
dc-tensors must satisfy conditions: $$
h_\alpha ^{\underline{\alpha }}(u,{\underline{u}})h_{\underline{\alpha }%
}^\beta ({\underline{u}},u)={\delta }_\alpha ^\beta ,h_\alpha ^{\underline{%
\alpha }}(u,{\underline{u}})h_{\underline{\beta }}^\alpha ({\underline{u}}%
,u)={\delta }_{\underline{\beta }}^{\underline{\alpha }} $$
and ${\lim }_{{({\underline{u}}{\to }u})}h_\alpha ^{\underline{\alpha }}(u,{%
\underline{u}})={\delta }_\alpha ^{\underline{\alpha }},\quad {\lim }_{{({%
\underline{u}}{\to }u})}h_{\underline{\alpha }}^\alpha ({\underline{u}},u)={%
\delta }_{\underline{\alpha }}^\alpha .$

Let ${\overline{S}}_p{\subset }{\overline{U}}{\subset }\overline{{V}}{{^{(n+m)}%
}}$ is a homeomorphic to $p$-dimensional sphere and suggest that
chains of na--maps are used to connect regions:
$$ U\ {\buildrel nc_{(1)} \over \longrightarrow}\ {\overline S}_p\ %
     {\buildrel nc_{(2)} \over \longrightarrow}\ {\underline U}.$$%

\begin{definition}
The tensor integral in ${\overline{u}}{\in }{\overline{S}}_p$ of a
dc--tensor $N_{{\varphi }{.}{\underline{\tau }}{.}{\overline{\alpha }}_1{%
\cdots }{\overline{\alpha }}_p}^{{.}{\gamma }{.}{\underline{\kappa }}}$ $({%
\overline{u}},u),$ completely antisymmetric on the indices ${{\overline{%
\alpha }}_1},{\ldots },{\overline{\alpha }}_p,$ over domain ${\overline{S}}%
_p,$ is defined as
\begin{eqnarray}
\label{tensint}N_{{\varphi }{.}{\underline{\tau }}}^{{.}{\gamma
}{.}{\underline{\kappa }}}({\underline{u}},u)
 &= &I_{({\overline{S}}_p)}^{\underline{U}}N_{{\varphi }{.}{\overline{\tau }}{.}
 {\overline{\alpha }}_1{\ldots }{\overline{\alpha }}_p}^{{.}{\gamma }{.}{\overline{\kappa }}}({\overline{u}},{%
\underline{u}})dS^{{\overline{\alpha }}_1{\ldots
}{\overline{\alpha }}_p}
  \\
& = &
{\int }_{({\overline{S}}_p)}h_{\underline{\tau }}^{\overline{\tau }}({%
\underline{u}},{\overline{u}})h_{\overline{\kappa }}^{\underline{\kappa }}({%
\overline{u}},{\underline{u}})N_{{\varphi }{.}{\overline{\tau
}}{.} {\overline{\alpha }}_1{\cdots }{\overline{\alpha
}}_p}^{{.}{\gamma }{.}{\overline{\kappa }}}
({\overline{u}},u)d{\overline{S}}^{{\overline{\alpha }}_1{\cdots
}{\overline{\alpha }}_p}, \nonumber
\end{eqnarray}
where $dS^{{\overline{\alpha }}_1{\cdots }{\overline{\alpha }}_p}={\delta }%
u^{{\overline{\alpha }}_1}{\land }{\cdots }{\land }{\delta }u_p^{\overline{%
\alpha }}$.
\end{definition}

Let suppose that transport dc--tensors $h_\alpha
^{\underline{\alpha }}$~ and $h_{\underline{\alpha }}^\alpha $~
admit covariant derivations of or\-der two and pos\-tu\-la\-te
ex\-is\-ten\-ce of de\-for\-ma\-ti\-on dc--ten\-sor $B_{{\alpha
}{\beta }}^{{..}{\gamma }}(u,{\underline{u}})$~ satisfying
relations
\begin{equation}
\label{rel1}D_\alpha h_\beta ^{\underline{\beta }}(u,{\underline{u}})=B_{{%
\alpha }{\beta }}^{{..}{\gamma }}(u,{\underline{u}})h_\gamma ^{\underline{%
\beta }}(u,{\underline{u}})
\end{equation}
and, taking into account that $D_\alpha {\delta }_\gamma ^\beta
=0,$ $$
D_\alpha h_{\underline{\beta }}^\beta ({\underline{u}},u)=-B_{{\alpha }{%
\gamma }}^{{..}{\beta }}(u,{\underline{u}})h_{\underline{\beta }}^\gamma ({%
\underline{u}},u). $$ By using the formulas for torsion and,
respectively, curvature of connection ${\Gamma }_{{\beta }{\gamma
}}^\alpha $~ we can calculate next commutators:
\begin{equation}
\label{com1}D_{[{\alpha }}D_{{\beta }]}h_\gamma ^{\underline{\gamma }}=-(R_{{%
\gamma }{.}{\alpha }{\beta }}^{{.}{\lambda }}+T_{{.}{\alpha
}{\beta }}^\tau B_{{\tau }{\gamma }}^{{..}{\lambda }})h_\lambda
^{\underline{\gamma }}.
\end{equation}
On the other hand from (\ref{rel1}) one follows that
\begin{equation}
\label{com2}D_{[{\alpha }}D_{{\beta }]}h_\gamma ^{\underline{\gamma }}=(D_{[{%
\alpha }}B_{{\beta }]{\gamma }}^{{..}{\lambda }}+B_{[{\alpha }{|}{\tau }{|}{.%
}}^{{..}{\lambda }}B_{{\beta }]{\gamma }{.}}^{{..}{\tau }})h_\lambda ^{%
\underline{\gamma }},
\end{equation}
where ${|}{\tau }{|}$~ denotes that index ${\tau }$~ is excluded
from the
action of antisymmetrization $[{\quad }]$. From (\ref{com1}) and (\ref{com2}%
) we obtain
\begin{equation}
\label{com3}D_{[{\alpha }}B_{{\beta }]{\gamma }{.}}^{{..}{\lambda }}+B_{[{%
\beta }{|}{\gamma }{|}}B_{{\alpha }]{\tau }}^{{..}{\lambda }}=(R_{{\gamma }{.%
}{\alpha }{\beta }}^{{.}{\lambda }}+T_{{.}{\alpha }{\beta }}^\tau B_{{\tau }{%
\gamma }}^{{..}{\lambda }}).
\end{equation}

Let ${\overline{S}}_p$~ be the boundary of ${\overline{S}}_{p-1}$.
The Stoke's type formula for the tensor--integral (\ref{tensint})
is defined as $$
I_{{\overline{S}}_p}N_{{\varphi }{.}{\overline{\tau }}{.}{\overline{\alpha }}%
_1{\ldots }{\overline{\alpha }}_p}^{{.}{\gamma }{.}{\overline{\kappa }}}dS^{{%
\overline{\alpha }}_1{\ldots }{\overline{\alpha }}_p}=I_{{\overline{S}}%
_{p+1}}{^{{\star }{(p)}}{\overline{D}}}_{[{\overline{\gamma }}{|}}N_{{%
\varphi }{.}{\overline{\tau }}{.}{|}{\overline{\alpha }}_1{\ldots }{{%
\overline{\alpha }}_p]}}^{{.}{\gamma }{.}{\overline{\kappa }}}dS^{{\overline{%
\gamma }}{\overline{\alpha }}_1{\ldots }{\overline{\alpha }}_p},
$$ where
\begin{eqnarray} &{}&
{^{{\star }{(p)}}D}_{[{\overline{\gamma }}{|}}N_{{\varphi
}{.}{\overline{\tau }}{.}{|} {\overline{\alpha }}_1{\ldots
}{\overline{\alpha }}_p]}^{{.}{\gamma }{.}{\overline{\kappa }}}=
D_{[{\overline{\gamma }}{|}}N_{{\varphi }{.}{\overline{\tau
}}{.}{|}{\overline{\alpha }}_1{\ldots } {\overline{\alpha
}}_p]}^{{.}{\gamma }{.}{\overline{\kappa }}} \nonumber \\
 &{}&
+pT_{{.}[{\overline{\gamma }}{\overline{\alpha
}}_1{|}}^{\underline{\epsilon }}N_{{\varphi }{.}{\overline{\tau
}}{.}{\overline{\epsilon }}{|}{\overline{\alpha }}_2{\ldots }
{\overline{\alpha }}_p]}^{{.}{\gamma }{.}{\overline{\kappa
}}}-B_{[{\overline{\gamma }}{|} {\overline{\tau
}}}^{{..}{\overline{\epsilon }}}N_{{\varphi
}{.}{\overline{\epsilon }}{.}{|} {\overline{\alpha }}_1{\ldots
}{\overline{\alpha }}_p]}^{{.}{\gamma }{.}{\overline{\kappa }}}+
B_{[{\overline{\gamma }}{|}{\overline{\epsilon
}}}^{..{\overline{\kappa }}}N_{{\varphi }{.}{\overline{\tau
}}{.}{|} {\overline{\alpha }}_1{\ldots }{\overline{\alpha
}}_p]}^{{.}{\gamma }{.}{\overline{\epsilon }}}. \nonumber
\end{eqnarray}
We define the dual element of the hypersurfaces element $dS^{{j}_1{\ldots }{j%
}_p}$ as
\begin{equation}
\label{hyperf}d{\cal S}_{{\beta }_1{\ldots }{\beta }_{q-p}}={\frac 1{{p!}}}{%
\epsilon }_{{\beta }_1{\ldots }{\beta }_{k-p}{\alpha }_1{\ldots }{\alpha }%
_p}dS^{{\alpha }_1{\ldots }{\alpha }_p},
\end{equation}
where ${\epsilon }_{{\gamma }_1{\ldots }{\gamma }_q}$ is
completely antisymmetric on its indices and $$
{\epsilon }_{12{\ldots }(n+m)}=\sqrt{{|}g{|}},g=det{|}g_{{\alpha }{\beta }{|}%
}, $$ $g_{{\alpha }{\beta }}$ is taken as the d--metric
(\ref{dmetric}). The dual
of dc--tensor $N_{{\varphi }{.}{\overline{\tau }}{.}{\overline{\alpha }}_1{%
\ldots }{\overline{\alpha }}_p}^{{.}{\gamma }{\overline{\kappa
}}}$ is
defined as the dc--tensor ${\cal N}_{{\varphi }{.}{\overline{\tau }}}^{{.}{%
\gamma }{.}{\overline{\kappa }}{\overline{\beta }}_1{\ldots }{\overline{%
\beta }}_{n+m-p}}$ satisfying
\begin{equation}
\label{dual1}N_{{\varphi }{.}{\overline{\tau }}{.}{\overline{\alpha }}_1{%
\ldots }{\overline{\alpha }}_p}^{{.}{\gamma }{.}{\overline{\kappa
}}}={\frac
1{{p!}}}{\cal N}_{{\varphi }{.}{\overline{\tau }}}^{{.}{\gamma }{.}{%
\overline{\kappa }}{\overline{\beta }}_1{\ldots }{\overline{\beta }}_{n+m-p}}%
{\epsilon }_{{\overline{\beta }}_1{\ldots }{\overline{\beta }}_{n+m-p}{%
\overline{\alpha }}_1{\ldots }{\overline{\alpha }}_p}.
\end{equation}
Using (\ref{hyperf}) and (\ref{dual1}) we can write
\begin{equation}
\label{dualint}I_{{\overline{S}}_p}N_{{\varphi }{.}{\overline{\tau }}{.}{%
\overline{\alpha }}_1{\ldots }{\overline{\alpha }}_p}^{{.}{\gamma }{.}{%
\overline{\kappa }}}dS^{{\overline{\alpha }}_1{\ldots }{\overline{\alpha }}%
_p}={\int }_{{\overline{S}}_{p+1}}{^{\overline{p}}D}_{\overline{\gamma }}%
{\cal N}_{{\varphi }{.}{\overline{\tau }}}^{{.}{\gamma
}{.}{\overline{\kappa
}}{\overline{\beta }}_1{\ldots }{\overline{\beta }}_{n+m-p-1}{\overline{%
\gamma }}}d{\cal S}_{{\overline{\beta }}_1{\ldots }{\overline{\beta }}%
_{n+m-p-1}},
\end{equation}
where
\begin{eqnarray} &{}& %
{^{\overline{p}}D}_{\overline{\gamma }}{\cal N}_{{\varphi }{.}%
{\overline{\tau }}}^{{.}{\gamma }{.}{\overline{\kappa }}{\overline{\beta%
}}_1{\ldots }{\overline{\beta }}_{n+m-p-1}{\overline{\gamma%
}}}={\overline{D}}_{\overline{\gamma }}{\cal N}_{{\varphi }{.}{\overline{\tau%
}}}^{{.}{\gamma }{.}{\overline{\kappa }}{\overline{\beta }}_1{\ldots%
}{\overline{\beta }}_{n+m-p-1}{\overline{\gamma }}}-B_{{\overline{\gamma%
}}{\overline{\tau }}}^{{..}{\overline{\epsilon }}}{\cal
N}_{{\varphi }{.}{\overline{\epsilon }}}^{{.}{\gamma
}{.}{\overline{\kappa }}{\overline{\beta }}_1{\ldots
}{\overline{\beta }}_{n+m-p-1}{\overline{\gamma }}} \nonumber \\
&{}&  +B_{{\overline{\gamma }}{\overline{\epsilon
}}}^{{..}{\overline{\kappa }}}%
{\cal N}_{{\varphi }{.}{\overline{\tau
}}}^{{.}{\gamma }{.}{\overline{\epsilon }}{\overline{\beta
}}_1{\ldots }{\overline{\beta }}_{n+m-p-1}{\overline{\gamma
}}}+(-1)^{(n+m-p)}(n+m-p+1)T_{{.}{\overline{\gamma
}}{\overline{\epsilon }}}^{[{\overline{\epsilon }}}{\cal
N}_{{\varphi }{.}{{\overline{\tau }}}}^{{.}{|}{\gamma
}{.}{\overline{\kappa }}{|}{\overline{\beta }}_1{\ldots
}{\overline{\beta }}_{n+m-p-1}]{\overline{\gamma }}}.   \nonumber
\end{eqnarray}  To verify the equivalence of (\ref{dual1}) and (\ref{dualint})
we must take  in consideration that   $$  D_\gamma {\epsilon
}_{{\alpha
}_1{\ldots }{\alpha }_k}=0\ \mbox{and}\ {%
\epsilon }_{{\beta }_1{\ldots
}{\beta }_{n+m-p}{\alpha }_1{\ldots }{\alpha }%
_p}{\epsilon }^{{\beta
}_1{\ldots }{\beta }_{n+m-p}{\gamma }_1{\ldots }{%
\gamma}_p}=p!(n+m-p)!{\delta }_{{\alpha }_1}^{[{\gamma }_1}{\cdots
}{\delta }_{{\alpha }_p}^{{\gamma }_p]}.   
$$  
The developed tensor integration formalism will be used in the next
section for definition of conservation laws on spaces with local
anisotropy.

\section{Tensor Integrals and Conservation Laws}

The definition of conservation laws on curved and/or
locally anisotropic  spaces is a challenging task because of
absence of global and local groups  of automorphisms of such
spaces. Our main idea is to use chains of na--maps  from a given,
called hereafter as the fundamental, la--spacetime to an
auxiliary one with trivial curvatures and torsions admitting a
global group  of automorphisms. The aim of this section is to
formulate conservation laws  for la-gravitational fields by using
dc--objects and tensor--integral  values, na--maps and variational
calculus on the Category of la--spacetimes.    R. Miron and M.
Anastasiei \cite{ma87,ma94} calculated the divergence of the
energy--momentum d--tensor on vector bundles provided with
N--connection  structure (the same formulas hold for (pseudo)
Riemannian la--spacetimes)   \begin{equation}
\label{divem}D_\alpha {E}_\beta ^\alpha ={\frac 1{\ {\kappa
}_1}}U_\alpha ,   \end{equation}  where   $$ {E}_\beta ^\alpha
=R_\beta ^\alpha -\frac 12\delta _\beta ^\alpha \overleftarrow{R}
$$ is the Einstein d--tensor, and concluded that the d--vector $$
U_\alpha ={\frac 12}(G^{\beta \delta }{{R_\delta }^\gamma }_{\phi \beta }{T}%
_{\cdot \alpha \gamma }^\phi -G^{\beta \delta }{{R_\delta }^\gamma
}_{\phi \alpha }{T}_{\cdot \beta \gamma }^\phi +{R_\phi ^\beta
}{T}_{\cdot \beta \alpha }^\phi ) $$
vanishes if and only if the d--connection $D$ is without torsion. On ${%
V^{(n+m)}}$ the d--torsion ${T}_{\cdot \alpha \gamma }^\phi $
could be effectively induced with respect to an anholonomic frame
and became trivial after transition to a holonomic frame.

No wonder that conservation laws, in usual physical theories being
a consequence of global (for usual gravity of local) automorphisms
of the fundamental spacetime, are more sophisticate on the spaces
with local anisotropy. Here it is important to emphasize the
multiconnection character of la--spacetimes. For example, for a
d--metric (\ref{dmetric}) on ${V^{(n+m)}} $ we can equivalently
define an auxiliar linear connection $\tilde D$ constructed from
by using the usual formulas for Christoffel symobls with the
operators of partial differential equations (\ref{dder}) chainged
respectively into the la--adapted ones (\ref{pder}). We conclude
that by using auxiliary symmetric d--connections, we can also use
the symmetric d--connection ${\gamma }_{\quad \beta \gamma
}^\alpha $ from (\ref{simdef}) we construct a model of la--gravity
which looks like locally isotropic on the spacetime ${V^{(n+m)}}.$
More general gravitational models with local anisotropy can be
obtained by using deformations of connection ${\tilde \Gamma
}_{\cdot \beta \gamma }^\alpha ,$ $$ {{\Gamma }^\alpha }_{\beta
\gamma }={\tilde \Gamma }_{\cdot \beta \gamma }^\alpha +{P^\alpha
}_{\beta \gamma }+{Q^\alpha }_{\beta \gamma }, $$ were, for
simplicity, ${{\Gamma }^\alpha }_{\beta \gamma }$ is chosen to be
also metric and satisfy the Einstein equations (\ref{einsteq2}).
The d--vector $U_\alpha $ is interpreted as an effective source of
local anisotropy on ${V^{(n+m)}}$ satisfying the generalized
conservation laws (\ref {divem}). The deformation d--tensor
${P^\alpha }_{\beta \gamma }$ is could be generated (or not) by
deformations of type (\ref{neqc1})--(\ref{neqc3}) for na--maps.

>From (\ref{tensint}) we obtain a tensor integral on ${\cal C}({V^{(n+m)}})$ of
a d--tensor: $$
N_{{\underline{\tau }}}^{{.}{\underline{\kappa }}}(\underline{u})=I_{{%
\overline{S}}_p}N_{{\overline{\tau }}{..}{\overline{\alpha }}_1{\ldots }{%
\overline{\alpha }}_p}^{{..}{\overline{\kappa }}}({\overline{u}})h_{{%
\underline{\tau }}}^{{\overline{\tau }}}({\underline{u}},{\overline{u}})h_{{%
\overline{\kappa }}}^{{\underline{\kappa }}}({\overline{u}},{\underline{u}}%
)dS^{{\overline{\alpha }}_1{\ldots }{\overline{\alpha }}_p}. $$

We point out that tensor--integrals can be defined not only for
dc--tensors but and for d--tensors on ${V^{(n+m)}}$. Really,
suppressing indices ${\varphi }$~ and ${\gamma }$~ in
(\ref{dual1}) and (\ref{dualint}), considering instead of a
deformation dc--tensor a deformation tensor
\begin{equation}
\label{deften1}B_{{\alpha }{\beta }}^{{..}{\gamma }}(u,{\underline{u}})=B_{{%
\alpha }{\beta }}^{{..}{\gamma }}(u)=P_{{.}{\alpha }{\beta
}}^\gamma (u)
\end{equation}
(we consider deformations induced by a nc--transform) and integration $%
I_{S_p}{\ldots }dS^{{\alpha }_1{\ldots }{\alpha }_p}$ in la--spacetime ${%
V^{(n+m)}}$ we obtain from (\ref{tensint}) a tensor--integral on ${\cal C}({%
V^{(n+m)}})$~ of a d--tensor: $$
N_{{\underline{\tau }}}^{{.}{\underline{\kappa }}}({\underline{u}}%
)=I_{S_p}N_{{\tau }{.}{\alpha }_1{\ldots }{\alpha }_p}^{.{\kappa }}(u)h_{{%
\underline{\tau }}}^\tau ({\underline{u}},u)h_\kappa ^{\underline{\kappa }%
}(u,{\underline{u}})dS^{{\alpha }_1{\ldots }{\alpha }_p}. $$
Taking into account (\ref{com2}) we can calculate that curvature
$$
{\underline{R}}_{{\gamma }{.}{\alpha }{\beta }}^{.{\lambda }}=D_{[{\beta }%
}B_{{\alpha }]{\gamma }}^{{..}{\lambda }}+B_{[{\alpha }{|}{\gamma }{|}}^{{..}%
{\tau }}B_{{\beta }]{\tau }}^{{..}{\lambda }}+T_{{.}{\alpha }{\beta }}^{{%
\tau }{..}}B_{{\tau }{\gamma }}^{{..}{\lambda }} $$
of connection ${\underline{\Gamma }}_{{.}{\alpha }{\beta }}^\gamma (u)={%
\Gamma }_{{.}{\alpha }{\beta }}^\gamma (u)+B_{{\alpha }{\beta }{.}}^{{..}{%
\gamma }}(u),$ with $B_{{\alpha }{\beta }}^{{..}{\gamma }}(u)$~ taken from (%
\ref{deften1}), vanishes, ${\underline{R}}_{{\gamma }{.}{\alpha }{\beta }}^{{%
.}{\lambda }}=0.$ So, we can conclude that a la--spacetime
${V^{(n+m)}}$ admits a tensor integral structure on ${\cal
{C}}({V^{(n+m)}})$ for d--tensors
associated to the deformation tensor $B_{{\alpha }{\beta }}^{{..}{\gamma }%
}(u)$ if the nc--image $\underline{{V}}{^{(n+m)}}$~ is locally
parallelizable. That way we generalize the one space tensor
integral constructions from \cite {goz,gv,v295a}, were the
possibility to introduce tensor integral structure on a curved
space was restricted by the condition that this space is locally
parallelizable. For $q=n+m$~ the relations (\ref{dualint}),
written for
d--tensor ${\cal N}_{\underline{\alpha }}^{{.}{\underline{\beta }}{%
\underline{\gamma }}}$ (we change indices ${\overline{\alpha }},{\overline{%
\beta }},{\ldots }$ into ${\underline{\alpha }},{\underline{\beta
}},{\ldots })$ extend the Gauss formula on ${\cal
{C}}({V^{(n+m)}})$:
\begin{equation}
\label{gausse}I_{S_{q-1}}{\cal N}_{\underline{\alpha }}^{{.}{\underline{%
\beta }}{\underline{\gamma }}}d{\cal S}_{\underline{\gamma }}=I_{{\underline{%
S}}_q}{^{\underline{q-1}}D}_{{\underline{\tau }}}{\cal N}_{{\underline{%
\alpha }}}^{{.}{\underline{\beta }}{\underline{\tau
}}}d{\underline{V}},
\end{equation}
where $d{\underline{V}}={\sqrt{{|}{\underline{g}}_{{\alpha }{\beta }}{|}}}d{%
\underline{u}}^1{\ldots }d{\underline{u}}^q$ and
\begin{equation}
\label{defrel}{^{\underline{q-1}}D}_{{\underline{\tau }}}{\cal N}_{%
\underline{\alpha }}^{{.}{\underline{\beta }}{\underline{\tau }}}=D_{{%
\underline{\tau }}}{\cal N}_{\underline{\alpha }}^{{.}{\underline{\beta }}{%
\underline{\tau }}}-T_{{.}{\underline{\tau }}{\underline{\epsilon }}}^{{%
\underline{\epsilon }}}{\cal N}_{{\underline{\alpha }}}^{{\underline{\beta }}%
{\underline{\tau }}}-B_{{\underline{\tau }}{\underline{\alpha }}}^{{..}{%
\underline{\epsilon }}}{\cal N}_{{\underline{\epsilon }}}^{{.}{\underline{%
\beta }}{\underline{\tau }}}+B_{{\underline{\tau }}{\underline{\epsilon }}}^{%
{..}{\underline{\beta }}}{\cal N}_{{\underline{\alpha }}}^{{.}{\underline{%
\epsilon }}{\underline{\tau }}}.
\end{equation}

Let consider physical values $N_{{\underline{\alpha }}}^{{.}{\underline{%
\beta }}}$ on $\underline{{V}}{{^{(n+m)}}}$~ defined on its density ${\cal N}_{%
{\underline{\alpha }}}^{{.}{\underline{\beta }}{\underline{\gamma
}}},$ i. e.
\begin{equation}
\label{density}N_{{\underline{\alpha }}}^{{.}{\underline{\beta }}}=I_{{%
\underline{S}}_{q-1}}{\cal N}_{{\underline{\alpha }}}^{{.}{\underline{\beta }%
}{\underline{\gamma }}}d{\cal S}_{{\underline{\gamma }}}
\end{equation}
with this conservation law (due to (\ref{gausse})):
\begin{equation}
\label{diverg2}{^{\underline{q-1}}D}_{{\underline{\gamma }}}{\cal N}_{{%
\underline{\alpha }}}^{{.}{\underline{\beta }}{\underline{\gamma
}}}=0.
\end{equation}
We note that these conservation laws differ from covariant
conservation laws for well known physical values such as density
of electric current or of
energy-- momentum tensor. For example, taking the density ${E}_\beta ^{{.}{%
\gamma }},$ with corresponding to (\ref{defrel}) and
(\ref{diverg2}) conservation law,
\begin{equation}
\label{diverg3}{^{\underline{q-1}}D}_{{\underline{\gamma }}}{E}_{{\underline{%
\beta }}}^{{\underline{\gamma }}}=D_{{\underline{\gamma }}}{E}_{{\underline{%
\beta }}}^{{\underline{\gamma }}}-T_{{.}{\underline{\epsilon }}{\underline{%
\tau }}}^{{\underline{\tau }}}{E}_{{\underline{\beta }}}^{{.}{\underline{%
\epsilon }}}-B_{{\underline{\tau }}{\underline{\beta }}}^{{..}{\underline{%
\epsilon }}}{E}_{\underline{\epsilon }}^{{\underline{\tau }}}=0,
\end{equation}
we can define values (see (\ref{gausse}) and (\ref{density})) $$
{\cal P}_\alpha =I_{{\underline{S}}_{q-1}}{E}_{{\underline{\alpha }}}^{{.}{%
\underline{\gamma }}}d{\cal S}_{{\underline{\gamma }}}. $$
The defined conservation laws (\ref{diverg3}) for ${E}_{{\underline{\beta }}%
}^{{.}{\underline{\epsilon }}}$ are not related with those for
energy--momentum tensor $E_\alpha ^{{.}{\gamma }}$ from the
Einstein
equations for the almost Hermitian gravity \cite{ma87,ma94} or with a ${%
\tilde E}_{\alpha \beta }$ with vanishing divergence $D_\gamma {\tilde E}%
_\alpha ^{{.}{\gamma }}=0.$ So ${\tilde E}_\alpha ^{{.}{\gamma }}{\neq }{E}%
_\alpha ^{{.}{\gamma }}.$ A similar conclusion was made in
\cite{goz} for the unispacial locally isotropic tensor integral.
In the case of multispatial tensor integration we have another
possibility (firstly pointed
in \cite{vog,v295a} for Einstein-Cartan spaces), namely, to identify ${E}_{{%
\underline{\beta }}}^{{.}{\underline{\gamma }}}$ from
(\ref{diverg3}) with the na-image of ${E}_\beta ^{{.}{\gamma }}$
on la--spacetime ${V^{(n+m)}}.$ We shall consider this
construction in the next Section.

\section{Conservation Laws for Na--Backrounds}

Let us consider a fixed background la--spacetime
$\underline{{V}}{{^{(n+m)}}}$
with given metric ${\underline{g}}_{\alpha \beta }=({\underline{g}}_{ij},{%
\underline{h}}_{ab})$ and d--connection ${\underline{\tilde \Gamma
}}_{\cdot \beta \gamma }^\alpha .$ For simplicity, we suppose that
the metricity conditions are satisfied and that the connection is
torsionless and with vanishing curvature. Considering a
nc--transform from the fundamental la--space ${V{^{(n+m)}}}$ to an
auxiliary one $\underline{{V}}{{^{(n+m)}}}$ we
are interested in the equivalents of the Einstein equations on $\underline{{V%
}}{{^{(n+m)}}}.$

We suppose that a part of gravitational degrees of freedom is
''pumped out''
into the dynamics of deformation d--tensors for d--connection, ${P^\alpha }%
_{\beta \gamma },$ and metric, $B^{\alpha \beta
}=(b^{ij},b^{ab}).$ The
remained part of degrees of freedom is coded into the metric ${\underline{g}}%
_{\alpha \beta }$ and d--connection ${\underline{\tilde \Gamma
}}_{\cdot \beta \gamma }^\alpha .$

Following \cite{gri,vrjp} we apply the first order formalism and consider $%
B^{\alpha \beta }$ and ${P^\alpha }_{\beta \gamma }$ as
independent variables on $\underline{{V}}{{^{(n+m)}}}.$ Using
notations $$
P_\alpha ={P^\beta }_{\beta \alpha },\quad {\Gamma }_\alpha ={{\Gamma }%
^\beta }_{\beta \alpha }, $$ $$ {\hat B}^{\alpha \beta
}=\sqrt{|g|}B^{\alpha \beta },\widehat{{g}}^{\alpha \beta
}=\sqrt{|g|}g^{\alpha \beta },{\underline{\widehat{g}}}^{\alpha
\beta }=\sqrt{|\underline{g}|}{\underline{g}}^{\alpha \beta } $$
and making identifications $$
{\hat B}^{\alpha \beta }+{\underline{\widehat{g}}}^{\alpha \beta }=\widehat{{%
g}}^{\alpha \beta },{\quad }{\underline{\Gamma }}_{\cdot \beta
\gamma }^\alpha -{P^\alpha }_{\beta \gamma }={{\Gamma }^\alpha
}_{\beta \gamma }, $$ we take the action of la--gravitational
field on $\underline{{V}}{{^{(n+m)}}}$ in this form:
\begin{equation}
\label{action3}{\underline{{\cal S}}}^{(g)}=-{(2c{\kappa }_1)}^{-1}\int {%
\delta }^qu{}{\underline{{\cal L}}}^{(g)},
\end{equation}
where $$ {\underline{{\cal L}}}^{(g)}={\hat B}^{\alpha \beta
}(D_\beta P_\alpha -D_\tau {P^\tau }_{\alpha \beta
})+({\underline{\widehat{g}}}^{\alpha \beta
}+{\hat B}^{\alpha \beta })(P_\tau {P^\tau }_{\alpha \beta }-{P^\alpha }%
_{\alpha \kappa }{P^\kappa }_{\beta \tau }) $$
and the interaction constant is taken ${\kappa }_1={\frac{4{\pi }}{{c^4}}}k,{%
\quad }(c$ is the light constant and $k$ is Newton constant) in
order to obtain concordance with the Einstein theory in the
locally isotropic limit.

We construct on $\underline{{V}}{{^{(n+m)}}}$ a la--gravitational
theory with matter fields (denoted as ${\varphi }_A$ with $A$
being a general index) interactions by postulating this Lagrangian
density for matter fields
\begin{equation}
\label{actmat1}{\underline{{\cal L}}}^{(m)}={\underline{{\cal L}}}^{(m)}[{%
\underline{\widehat{g}}}^{\alpha \beta }+{\hat B}^{\alpha \beta
};{\frac
\delta {\delta u^\gamma }}({\underline{\widehat{g}}}^{\alpha \beta }+{\hat B}%
^{\alpha \beta });{\varphi }_A;{\frac{\delta {\varphi }_A}{\delta u^\tau }}%
].
\end{equation}

Starting from (\ref{action3}) and (\ref{actmat1}) the total action
of la--gravity on $\underline{{V}}{{^{(n+m)}}}$ is written as
\begin{equation}
\label{actiontotal}{\underline{{\cal S}}}={(2c{\kappa
}_1)}^{-1}\int {\delta
}^qu{\underline{{\cal L}}}^{(g)}+c^{-1}\int {\delta }^{(m)}{\underline{{\cal %
L}}}^{(m)}.
\end{equation}
Applying variational procedure on $\underline{{V}}{{^{(n+m)}}},$
similar to that presented in \cite{gri} but in our case adapted to
N--connection by using derivations (\ref{dder}) instead of partial
derivations (\ref{pder}), we derive from (\ref{actiontotal}) the
la--gravitational field equations
\begin{equation}
\label{lagraveq}{\bf {\Theta }}_{\alpha \beta }={{\kappa }_1}({\underline{%
{\bf t}}}_{\alpha \beta }+{\underline{{\bf T}}}_{\alpha \beta })
\end{equation}
and matter field equations
\begin{equation}
\label{mattereq}{\frac{{\triangle }{\underline{{\cal L}}}^{(m)}}{\triangle {%
\varphi }_A}}=0,
\end{equation}
where $\triangle /\triangle {\varphi }_A$ denotes the variational
derivation.

In (\ref{lagraveq}) we have introduced these values: the
energy--momentum d--tensor for la--gravi\-ta\-ti\-on\-al field $$
{\kappa }_1{\underline{{\bf t}}}_{\alpha \beta }=({\sqrt{|g|}})^{-1}{\frac{%
\triangle {\underline{{\cal L}}}^{(g)}}{\triangle g^{\alpha \beta }}}%
=K_{\alpha \beta }+{P^\gamma }_{\alpha \beta }P_\gamma -{P^\gamma
}_{\alpha \tau }{P^\tau }_{\beta \gamma }+ $$
\begin{equation}
\label{energmom1}{\frac 12}{\underline{g}}_{\alpha \beta }{\underline{g}}%
^{\gamma \tau }({P^\phi }_{\gamma \tau }P_\phi -{P^\phi }_{\gamma \epsilon }{%
P^\epsilon }_{\phi \tau }),
\end{equation}
(where $$ K_{\alpha \beta }={\underline{D}}_\gamma K_{\alpha \beta
}^\gamma , $$ $$
2K_{\alpha \beta }^\gamma =-B^{\tau \gamma }{P^\epsilon }_{\tau (\alpha }{%
\underline{g}}_{\beta )\epsilon }-B^{\tau \epsilon }{P^\gamma
}_{\epsilon (\alpha }{\underline{g}}_{\beta )\tau }+ $$ $$
{\underline{g}}^{\gamma \epsilon }h_{\epsilon (\alpha }P_{\beta )}+{%
\underline{g}}^{\gamma \tau }{\underline{g}}^{\epsilon \phi }{P^\varphi }%
_{\phi \tau }{\underline{G}}_{\varphi (\alpha }B_{\beta )\epsilon }+{%
\underline{g}}_{\alpha \beta }B^{\tau \epsilon }{P^\gamma }_{\tau
\epsilon }-B_{\alpha \beta }P^\gamma {\quad }), $$ $$
2{\bf \Theta }={\underline{D}}^\tau {\underline{D}}_\tau B_{\alpha \beta }+{%
\underline{g}}_{\alpha \beta }{\underline{D}}^\tau
{\underline{D}}^\epsilon
B_{\tau \epsilon }-{\underline{g}}^{\tau \epsilon }{\underline{D}}_\epsilon {%
\underline{D}}_{(\alpha }B_{\beta )\tau } $$ and the
energy--momentum d--tensor of matter
\begin{equation}
\label{energmomt}{\underline{{\bf T}}}_{\alpha \beta
}=2{\frac{\triangle
{\cal L}^{(m)}}{\triangle {\underline{\widehat{g}}}^{\alpha \beta }}}-{%
\underline{g}}_{\alpha \beta }{\underline{g}}^{\gamma \delta }{\frac{%
\triangle {\cal L}^{(m)}}{\triangle
{\underline{\widehat{g}}}^{\gamma \delta }}}.
\end{equation}
As a consequence of (\ref{mattereq})--(\ref{energmomt}) we obtain
the d--covariant on $\underline{{V}}{{^{(n+m)}}}$ conservation
laws
\begin{equation}
\label{dcovar}{\underline{D}}_\alpha ({\underline{{\bf t}}}^{\alpha \beta }+{%
\underline{{\bf T}}}^{\alpha \beta })=0.
\end{equation}
We have postulated the Lagrangian density of matter fields
(\ref{actmat1})
in a form as to treat ${\underline{{\bf t}}}^{\alpha \beta }+{\underline{%
{\bf T}}}^{\alpha \beta }$ as the source in (\ref{lagraveq}).

Now we formulate the main results of this Section:

\begin{proposition}
The dynamics of the  la--gravitational fields, modeled as
solutions of the Einstein equations (\ref{einsteq2}) and of matter
field equations on la--spacetime ${V{^{(n+m)}}},$ can be locally
equivalently modeled on a background la--spacetime
$\underline{{V}}{{^{(n+m)}}}$ provided with a trivial d-connection
and metric structure (with vanishing d--tensors of torsion and
curvature) by  equations (\ref{lagraveq}) and (\ref{mattereq}) on
condition that the deformation tensor ${P^\alpha }_{\beta \gamma
}$ is a solution of the Cauchy problem posed for the basic
equations for a chain of na--maps from ${V{^{(n+m)}}}$ to
$\underline{{V}}{{^{(n+m)}}}.$
\end{proposition}

\begin{proposition}
The local  d--tensor  conservation laws for Einstein
la--gravita\-ti\-on\-al
fields can be written in the form (\ref{dcovar}) for both la--gravita\-ti\-on\-al (%
\ref{energmom1}) and matter (\ref{energmomt}) energy--momentum
d--tensors.
These laws are d--covariant on the background space $\underline{{V}}{{^{(n+m)}}%
}$ and must be completed with invariant conditions of type (\ref{nainv0})--((%
\ref{nainv3})) for every deformation parameters of a chain of na--maps from $%
{V{^{(n+m)}}}$ to $\underline{{V}}{{^{(n+m)}}}.$
\end{proposition}

The above presented considerations consist proofs of both
propositions.

We emphasize that the nonlocalization of both locally
an\-i\-sot\-rop\-ic and isot\-rop\-ic gravitational
energy--momentum values on the fundamental (locally
an\-i\-sot\-rop\-ic or isotropic) spacetime ${V{^{(n+m)}}}$ is a
consequence of the absence of global group automorphisms for
generic curved spaces. Considering gravitational theories from the
view of multispaces and their mutual maps (directed by the basic
geometric structures on ${V{^{(n+m)}}} $ such as N--connection,
d--connection, d--torsion and d--curvature components, see the
coefficients for basic na--equations (\ref{neqc1})--(\ref
{neqc3})), we can formulate local d--tensor conservation laws on
auxiliary globally automorphic spaces being related with some
covering regions of the spacetime ${V{^{(n+m)}}}$ by means of
chains of na--maps. Finally, we remark that as a matter of
principle we can also use d--connection deformations in order to
modelate the la--gravitational interactions with nonvanishing
torsion and nonmetricity. In this case we must introduce a
corresponding source in (\ref {dcovar}) and define generalized
conservation laws as in (\ref{divem}).

\section{Einstein Spaces Generated  by Finsler Like Metrics}

In this Section we analyze the conditions when  four dimensional
(4D) vacuum and non--vacuum solutions of the Einstein equations
can be induced by  Finsler like metrics depending on three
variables; we construct such solutions in explicit form.
\subsection{Two dimensional Finsler metrics}
There is a class of 2D Finsler metrics
\begin{equation}
\label{fm1}h_{ab}\left( x^i,y^c\right) =\frac 12\frac{\partial
^2F^2\left( x^i,y^c\right) }{\partial y^a\partial y^b}
\end{equation}
generated by the so--called Finsler metric function $F=F\left(
x^i,y^c\right) ,$ where the indices $i,j,k,...\,$ run values $1$
and $2$ on a 2D base manifold $V^{(2)}$ and $y$--coordinate indices $%
a,b,c,...=3,4$ are used for 2D fibers $Y_x$ of the tangent bundle $TV^{(2)}$%
.. Because for Finsler spaces the function $F$ is homogeneous on $y$%
--variables we can express
\begin{equation}
\label{finm1}F\left( x^1,x^2,y^3,y^4\right) =y^3f(x^1,x^2,z),
\end{equation}
where $z=y^4/y^3$ and $$ f=f(x,z)=f(x^1,x^2,z)\doteq F\left(
x^1,x^2,1,z\right) . $$ By introducing the function $K\left(
x,z\right) =\left( ff^{\prime }\right) ^{\prime },$ where the
'prime' denotes the partial derivation on $z,$ the metric
coefficients (\ref{fm1}) are computed
\begin{equation}
\label{projm}h_3 = h_{33}=Kz^2-2ff^{\prime }z+f^2, h=
h_{34}=-Kz+ff^{\prime}, h_4 = h_{44}=K.
\end{equation}
We note that if the 2D Finsler metric coefficients formally
depended on four variables, by introducing the function
$f(x^1,x^2,z)$ one has obtained an explicit dependence only on
three coordinates.

Consider a vertical 2D d-metric
\begin{equation}
\label{ndiag}h_{ab}(x^i,z)=\left(
\begin{array}{cc}
h_3(x^i,z) & h(x^i,z) \\ h(x^i,z) & h_4(x^i,z)
\end{array}
\right)
\end{equation}
which, by applying a matrix transform (\ref{diagonalization}) can
be diagonalized
\begin{equation}
\label{diag2}h_{a^{\prime }b^{\prime }}(x^i,z)=\left(
\begin{array}{cc}
\lambda _{3}(x^i,z) & 0 \\ 0 & \lambda _{4}(x^i,z)
\end{array}
\right).
\end{equation}
 We can generate by a Finsler metric function a 2D
diagonal (pseudo) Riemannian metric
\begin{equation}
\label{diagnz}h_{a^{\prime }b^{\prime }}(t,r)=\left(
\begin{array}{cc}
\lambda _3(x^i) & 0 \\ 0 & \lambda _4(x^i)
\end{array}
\right)
\end{equation}
depending only on coordinates $x=\{x^i\}$ if we choose the square of $f$%
--function $s(x^i,z)=f^2\left( x^i,z\right) $ from (\ref{finm1})
to be (see formulas (\ref{projm}))
\begin{equation}
\label{rfm}s(x^i,z)=z^2\lambda _2(x^i)+\lambda _1(x^i).
\end{equation}
This is the simplest case when a 2D diagonal metric (\ref{diagnz})
is
defined by a trivial Finsler squared $f$--function (\ref{rfm}) depending on $%
z^2$ and two functions $\lambda _{1,2}(x^i).$

The problem of definition of a corresponding Finsler metric
function becames more difficult if we try to generate not a
diagonal 2D metric (\ref{diagnz}) depending only on two variables
$(x^i),$ but a nondiagonal one depending on three variables
$(x^i,z)$ (see (\ref{ndiag})). There are three classes of such
type Finsler generated 2D metrics.

\subsubsection{Euler nonhomogeneous equations and Finsler metrics}

The first class of 2D Finsler metric is defined by the condition
when the function $s(x^i,z)$ is chosen as to solve the first
equation in (\ref{projm})
 when the coefficient $a_1(x^i,z)$ of a nondiagonal 2D d--metric (\ref
{ndiag}) are prescribed. The rest of components of the vertical d--metric, $%
b_1(x^i,z)$ and $h(x^i,z),$ are not arbitrary ones but they must
be found by using partial derivatives $s^{\prime }=\partial
s/\partial z$ and $s^{\prime \prime }=\partial ^2s/\partial z^2,$
in correspondence with the formulas (\ref {projm}).

The basic equation is
\begin{equation}
\label{euler}z^2s^{\prime \prime }-2zs^{\prime }+2s=2a_1
\end{equation}
which for $a_1=0$ and variables $x^i$ treated as some parameters
is the so--called Euler equation \cite{kamke} having solutions of
type $$ C_1(x^i)z^2+C_2(x^i)z. $$
By integrating  on the $z$%
--variable we can construct the solution $s_{(a1)}$ of
(\ref{euler}) for a nonvanishing right part,
\begin{equation}
\label{seuler}s_{(a1)}\left( x^i,z\right) =
zC_{(0)}\left(x^i\right) + z^2C_{(1)}\left(x^i\right) + z\cdot
\int\limits_{const2}^zd\zeta \int\limits_{const1}^\zeta d\tau
\frac{a_1(x^i,\tau )}{\tau ^3},
\end{equation}
where $C_{(0)}\left( x^i\right) $ and $C_{(1)}\left( x^i\right) $
are some arbitrary functions, the index $(a1) $ emphasizes that
the Finsler metric is associated to the value $a_1(x^i,\tau )$ and
the $const1$ and $const2$ in the integrals should be chosen from
some boundary conditions.

The 2D d-metric coefficients $h(x^i,\tau )$ and $b_1(x^i,\tau )$
are computed $$ b_1=\frac 12s^{\prime \prime }\mbox{ and
}h=-zb_1+\frac{s^{\prime }}2, $$ where $s=s_{(a_1)}.$

\subsubsection{The simplest case}

If the coefficient $b_1(x^i,\tau )$ is given, the squared 2D
metric Finsler
function is to be found from the last formula in (\ref{projm}), $%
b_1=s^{\prime \prime }/2.$ By considering the coordinates $\left(
r,t\right) $ as parameters, and integrating on $z$ we obtain $$
s_{(b)}=2\int\limits_{const1}^zd\tau \int\limits_{const2}^\tau
d\vartheta b_1\left( x^i,\vartheta \right) +zS_{(0)}\left(
x^i\right) +S_{(1)}\left( x^i\right) , $$ where $S_{(0)}\left(
x^i\right) $ and $S_{(1)}\left( x^i\right) $ are some functions on
variables $x^i.$

The corresponding 2D vertical d--metric coefficients $h(x^i,\tau )$ and $%
a_1(x^i,\tau )$ are computed%
$$ a_1=s-z^2b_1-2zh\mbox{ and }h=-zb_1+\frac{s^{\prime }}2, $$
where $s=s_{(b).}$

\subsubsection{Prescribed nondiagonal coefficients}

In this case one choose the coefficient $h(x^i,\tau )$ for
definition of the squared Finsler metric function $s(x^i,\tau ).$
As the basic equation we  consider the equation $$ zs^{\prime
\prime }-s^{\prime }=-h $$ which has the solution $$
s_{(h)}=\varphi _1\left( x^i\right) +z^2\varphi _2\left(
x^i\right) -2\int\limits_{const1}^z\zeta d\zeta \cdot
\int\limits_{const2}^\zeta d\tau \frac{h(x^i,\tau )}{\tau ^2} $$
depending on two arbitrary functions $\varphi _{1,2}\left(
t,r\right) .$

The explicit formulas for the rest of 2D vertical d--metric coefficients $%
b_1(x^i,\tau )$ and $a_1(x^i,\tau )$ follows from $$
h=-zb_1+\frac{s^{\prime }}2\mbox{ and }a_1=s-z^2b_1-2zh $$ where
$s=s_{(h)}.$

\subsection{An ansatz for Finsler like vacuum solutions}

Let us  consider a particular case of  metrics (\ref{ansatz}) are
generated as  generalized Lagrange metrics (\ref{glm}) by a
diagonalization transform (\ref{diagonalization}) of a Finsler
induced v--metric (\ref{diag2})
\begin{equation}
\left[
\begin{array}{cccc}
g_1+q_1{}^2h_3+n_1{}^2\lambda _4 & 0 & q_1 \lambda _3 & n_1
\lambda _4 \\ 0 & 1+q_2{}^2\lambda _3+n_2{}^2\lambda _4 &
q_2\lambda _3 & n_2\lambda _4 \\ q_1\lambda _3 & q_2\lambda _3 &
\lambda _3 & 0 \\ n_1\lambda _4 & n_2\lambda _4 & 0 & \lambda _4
\end{array}
\right]   \label{ansatz2}
\end{equation}
with coefficients being some functions of necessary smo\-oth class $%
g_1=g_1(x^2), g_2= 1, q_i=q_i(x^j,z),n_i=n_i(x^j,z),$ $\lambda _3
= \lambda _3(x^j,z)$ and $\lambda _4 = \lambda _4(x^j).$ Latin
indices run respectively $i,j,k,...$ $=1,2$ and $a,b,c,...=3,4$
and the local coordinates are denoted $u^\alpha =(x^i,y^3=z,y^4),$
where one from the coordinates $x^1, z$ and $y^4$
 could be treated as a timelike coordinate. A metric  (\ref{ansatz2})  is diagonalized,
\begin{equation} \label{diag}
\delta s^2=g_1(x^2)\left( dx^1\right) ^2+ \left( dx^2\right) ^2
+\lambda _a (x^j,z)\left( \delta y^a\right)^2,
\end{equation}
with respect to corresponding anholonomic frames (\ref{dder}) and
(\ref{ddif}), here we write down only the 'elongated'
differentials $$ \delta z=dz+q_i(x^j,z)dx^i,\ \delta
y^4=dy^4+n_i(x^j,z)dx^i. $$

The system of Einstein field equations (\ref{einsteq2}) reduces to
four nontrivial second order partial differential
 equations on $z$ for functions $q_i(x^j,z), n_i(x^j,z), \lambda _3(x^j,z)$ and $\lambda _4(x^j),$
 \begin{eqnarray} \label{vac1}
  P_{3i} &=& {\frac{q_i}{2 \lambda _3}} \lbrack {\frac{1}{\lambda _3}}
  {\left( {\frac{\partial \lambda _3}{\partial z}}\right)}^2 -
 {\frac{{\partial}^2 \lambda _3}{\partial z^2}}\rbrack ,  \\
  \label{vac2}
  P_{4i} &= &{\frac{\lambda _4}{4 \lambda _3}} \lbrack {\frac{\partial n_i}{\partial z}}
   {\frac{\partial \lambda _3}{\partial z}} - 2 {\frac{{\partial}^2 n_i}{\partial z^2}}\rbrack .
 \end{eqnarray}

 There are two possibilities to satisfy the equations (\ref{vac1}): 1/ If the function $\lambda _3(x^j,z)$
 is a nonvanishing solution of
 \begin{equation} {\frac{1}{\lambda _3}} {\left( {\frac{\partial \lambda _3}{\partial z}}\right)}^2 -
 {\frac{{\partial}^2 \lambda _3}{\partial z^2}} = 0, \label{condition1}
 \end{equation}
 the N--connection coefficients $q_i(x^j,z)$ could take arbitrary values in correspondence to a stated Cauchy
 problem. 2/ The coefficients $q_i(x^j,z)\equiv 0$ if the function ${\lambda _3}$ das not satisfy the condition
  (\ref{condition1}); we have only one anisotropic direction distinguished by some nontrivial functions
  $n_i(x^j,z).$

  The general solution of (\ref{vac2}) is written
  \begin{equation} \label{condition2}
  n_i(x^j,z)=p_{i(0)}(x^i)\int_0^z \exp{\lbrack \frac{{\lambda}_3(x^j,\zeta)}{2} \rbrack} d\zeta + n_{i(0)}(x^i),
  \end{equation}
  where the functions $p_{i(0)}(x^i)$ and $n_{i(0)}(x^i)$ have to be defined from some
   boundary (initial) conditions.

  We conclude this Section by formulating the rule for generation by  Finsler like metrics of
    vacuum solutions of the Einstein equations. Firstly, we take a Finsler metric function (\ref{finm1}) and
   following  (\ref{projm}) we induce a nondiagonal 2D v--metric (\ref{ndiag}). Diagonalizing
   (\ref{diagonalization}),  we obtain a h--metric (\ref{diag2}). If the induced coefficient ${\lambda}_4$ depends
    only on horizontal variables $x^j,$
     the ansatz (\ref{ansatz2}) solves the vacuum Einsten  equations under the conditions that the functions
  $q_i(x^j,z), n_i(x^j,z),$ $\lambda _3(x^j,z)$ and $\lambda _4(x^j)$ satisfy the conditions (\ref{condition1})
   and (\ref{condition2}). Instead of starting the procedure by fixing the Finsler metric function we can
   fix a necessary type coeffcient $h_3$ ($h$ or $h_4$) and then, as was stated in subsection 8.1.1 (8.1.2 or 8.1.3),
    we must define the corresponding class of Finsler metrics. Finally,  diagonalizing  the v--metric,
     we obtain the coefficients which must be put into the ansatz (\ref{ansatz2}).

     We restricted our constructions only for some trivial Finsler like induced h--components of d--metrics, for
      simplicity, considering h-components of type $g_{ij} = diag[a_1(x^2),1].$ To induce more general Finsler
      like h--metrics is possible by a similar to the presented for  v--subspaces procedure.

\subsection{Non vacuum locally anisotropic solutions}

In this subsection we generalize the ansatz in order to induce
non--vacuum solutions of the Einstein equations (\ref{ansatz2}).
We consider a 4D metric parametrized
\begin{equation}
\left[
\begin{array}{cccc}
g_1+q_1{}^2h_3+n_1{}^2\lambda _4 & 0 & q_1 \lambda _3 & n_1
\lambda _4 \\ 0 & g_2+q_2{}^2\lambda _3+n_2{}^2\lambda _4 &
q_2\lambda _3 & n_2\lambda _4 \\ q_1\lambda _3 & q_2\lambda _3 &
\lambda _3 & 0 \\ n_1\lambda _4 & n_2\lambda _4 & 0 & \lambda _4
\end{array}
\right]   \label{ansatz3}
\end{equation}
with the coefficients being some functions of necessary smo\-oth class $%
g_1=-\alpha (r), g_2= 1/\alpha (r),
q_i=q_i(x^j,z),n_i=n_i(x^j,z),$ $\lambda _3 = \lambda _3(x^j,z)$
and $\lambda _4 = \lambda _4(x^j,z)$ where the h--coordinates are
denoted $x^1=t$
 (the time like coordinate) and $x^2=r.$  Our aim is to
 define the function $\alpha (r)$ which gives a solution of the Einstein equations with diagonal
 energy momentum d--tensor $$\Upsilon ^\beta _\gamma =[-\varepsilon, p_2, p_3, p_4]$$ for a matter state when
 $p_2=-\varepsilon$ and $p_3 = p_4.$

Putting these values of h--metric into (\ref{dricci}) we compute
$$ R_1^1=R_2^2=-\frac 12 {\ddots \alpha}, $$ where the dot denote
the partial derivative on $r.$
Considering the 2D h--subspace to be of constant negative scalar curvature,%
$$ \widehat{R}=2R_1^1=-\widetilde{m}^2, $$ and that the Einstein
la--equations (\ref{einsteq2}) are satisfied we obtain the
relation
\begin{equation}
\label{bheq1} {\ddots \alpha} = \widetilde{m}^2=\kappa \Upsilon
_3^3=\kappa \Upsilon _4^4.
\end{equation}

The solution of (\ref{bheq1}) is written in the form $\alpha
=\left( \widetilde{m}^2r^2-M\right) $ which defines a 2D h--metric
\begin{equation}
\label{bh1}ds_{(h)}^2=-\left( \widetilde{m}^2r^2-M\right)
dt^2+\left( \widetilde{m}^2r^2-M\right) ^{-1}dr^2
\end{equation}
being similar to a black hole solution in 2D Jackiw--Teitelboim
gravity \cite{jackiw} and display many of attributes of black
holes \cite
{mann,gegenberg1,lemos1} with that difference that the constant $\widetilde{m%
}$ is defined by 4D physical values in v--subspace and for
definiteness of the theory the h--metric should be supplied with
the equations for
 the  v--components of the d--metric which in our case is
 $$ \frac{\partial ^2 \lambda _4}{\partial z^2}-
 {\frac{1}{2\lambda _4}} \left( \frac{\partial \lambda _4}{\partial z}  \right)^2
  - {\frac{1}{2\lambda _3}}\left( \frac{\partial \lambda _3}{\partial z}  \right)
  \left( \frac{\partial \lambda _4}{\partial z}  \right) +
  {\frac{\kappa \varepsilon}{2}} \lambda _3  \lambda _4 = 0.$$
  Prescribing one of the functions $\lambda _3,$ or $\lambda _4,$
  the second one is to be defined  by integration on the $z$--variable (see detailes in
  \cite{v2000}). One has a forth order partial differential equation for the
   metric function $f(x^i,z),$  see (\ref{finm1}) if we try to induce the horizontal part
   in a pure Finsler like manner.

\section{Nearly Conformally  Flat Gravitational Fields}

We analyze chains of na--maps which by corresponding deformation
parameters and deformations of connections induce a vacuum
d--metric  (\ref{diag}) (for non--vacuum metrics considerations
are similar).

The nontrivial canonical d--connection coefficients (\ref{cdcon})
 and d--torsions (\ref{dtorsions})  are respectively computed
\begin{eqnarray}
L^1_{\ 12} &=& L^1_{\ 21}= {\frac{1}{2}} {\frac{\partial \ln{|a_1
|}}{\partial x^2}},\ L^2_{\ 22} = - {\frac{1}{2}}{\frac{\partial
a_1 }{\partial x^2}};\ C^3_{\ 33} = {\frac{1}{2}}{\frac{\partial
\ln{|\lambda _3 |}}{\partial z}}; \label{cdconm} \\
\nonumber L^3_{\ 3i} &=&  {\frac{1}{2\lambda _3}} \left(
{\frac{\partial \lambda _3 }{\partial x^i}} -
 q_i {\frac{\partial h_3 }{\partial z}} \right),\
 L^3_{\ 4i} = - {\frac{\lambda _4}{2\lambda _3}} {\frac{\partial n_i}{\partial z}},\
 L^4_{\ 3i} = {\frac{1}{2}}{\frac{\partial n_i}{\partial z}},\
 L^4_{\ 4i} = {\frac{1}{2}}{\frac{\partial \ln{|\lambda _4 |}}{\partial x^i}},
\end{eqnarray}
and
\begin{eqnarray}
T^3_{\ 3i} &=& - T^3_{\ i3}= {\frac{\partial q_i}{\partial z}} -
L^3_{\ 3i},\ T^3_{\ 4i} = - T^3_{\ i4} = - L^3_{\ 4i},
\label{dtorsionsm}  \\ T^4_{\ 3i} &=& - T^4_{\ i3}=
{\frac{\partial n_i}{\partial z}} - L^4_{\ 3i},\ T^4_{\ 4i} = -
T^4_{\ i4} = - L^4_{\ 4i},  \nonumber  \\ T^3_{\ 12} &=& - T^3_{\
21}= {\frac{\partial q_1}{\partial x^2}} - {\frac{\partial
q_2}{\partial x^1}}
 - q_2 {\frac{\partial q_1}{\partial z}}+ q_1 {\frac{\partial q_2}{\partial z}},\nonumber \\
T^4_{\ 12} &=& - T^4_{\ 21}= {\frac{\partial n_1}{\partial x^2}} -
{\frac{\partial n_2}{\partial x^1}}
 - q_2 {\frac{\partial n_1}{\partial z}}+ q_1 {\frac{\partial n_2}{\partial z}}.\nonumber
\end{eqnarray}
The obtained values allow us to define some na--map chains, for
instance, from the Minkowski
 spacetime $V_{[0]}=M^{3,1},$ where it is pointed the spacetime signature (3,1),
  to a curved one with local anisotropy, $V^{(2+2)},$ provided by a metric (\ref{ansatz2}) (equvalently,
  a d--metric (\ref{diag})).

 In this Section we shall consider sets of invertible na--maps (the inversce to a na--transform
 is also considered to be a na--map) when we could neglect quadratic terms like
 $ P P $ and $ F F$ \cite{sin} in the basic na--equations (\ref{neqc1}), (\ref{neqc2}) and
 (\ref{neqc3}), for simplicity, taken in a nonsymetrized form:
 \begin{itemize}
 \item for $na_{(1)}$--maps
  \begin{equation}
\label{neqc1a}D_{{\alpha }}P_{{.}{\beta }{\gamma }}^\delta -
Q_{{.}{\alpha } {\tau }}^\delta P_{{\beta }{\gamma }}^\tau
 =b_{{\alpha }}P_{{.}{\beta }{\gamma }}^{{\delta }}
 +a_{{\beta }{\gamma }}{\delta }_{{\alpha }}^\delta ;
\end{equation}
\item  for $na_{(2)}$--maps the deformation d--tensor is parametrized
\begin{equation}
P_{{.}{\alpha }{\beta }}^\tau (u) = {\sigma }_{({\alpha
}}F_{{\beta })}^\tau  \label{defcon2a}
\end{equation}
and the basic  equations are taken
\begin{equation}
\label{neqc2a}{D}_{{\gamma }}F_{{\beta }}^\alpha - Q_{{.}{\tau }{\gamma }}^\alpha F_{{%
\beta }}^\tau ={\mu }_{{\gamma }}F_{{\beta }}^\alpha +{\nu }_{{\gamma }}{%
\delta }_{{\beta }}^\alpha;
\end{equation}
\item  for $na_{(3)}$--maps the deformation d--tensor is parametrized
\begin{equation}
P_{{.}{\beta }{\gamma }}^\alpha (u) = {\sigma }_{{\beta }{\gamma
}}{\varphi }^\alpha , \label{defcon3a}
\end{equation}
and the basic equations
\begin{equation}
\label{neqc3a}D_\beta {\varphi }^\alpha ={\nu }{\delta }_\beta ^\alpha +{\mu }%
_\beta {\varphi }^\alpha +{\varphi }^\gamma Q_{{.}{\gamma }{\delta
}}^\alpha.
\end{equation}
\end{itemize}

\subsection{Chains of $na_{(1)}$--maps}

We illustrate that the canonical d--connections and d--torsions of
the mentioned vacuum metrics with local anisotropy could be
induced by a chain of three $na_{(1)}$--transforms
\begin{equation}
{V_{[0]}}\ {\buildrel {na_{1}^{[1]}} \over \longrightarrow}\  {V_{[1]}}\ %
{\buildrel {na_{1}^{[2]}} \over \longrightarrow}\ {V_{[2]}}\
{\buildrel {na_{1}^{[3]}} \over \longrightarrow}\ {V^{(2+2)}}.
\label{chain_n1}
\end{equation}

The first step in this chain is defined by some deformations of
the symmetric (\ref{simdef})
 and antisymmetric (\ref{torsdef}) parts of the d--connection
\begin{eqnarray}
{na_{1}^{[1]}}: &{}& ^{[1]}\gamma _{\beta \tau}^\alpha =
 ^{[0]}\gamma _{\beta \tau}^\alpha + ^{[1]}P_{\beta \tau}^\alpha, \nonumber \\
 &{}& ^{[1]}T _{\beta \tau}^\alpha =
 ^{[0]}T _{\beta \tau}^\alpha + ^{[1]}Q_{\beta \tau}^\alpha, \nonumber
\end{eqnarray}
where it is considered that on the flat background space is chosen
a system of coordinates for which $^{[0]}\gamma _{\beta
\tau}^\alpha =  0$ and $^{[0]}T _{\beta \tau}^\alpha = 0. $ The
values $$( ^{[1]}b_\alpha = \partial _\alpha \ln{|L^1_{\ 12}|},
 ^{[1]}a_{\beta \gamma} =0,\ ^{[1]}P_{\beta \gamma}^\alpha = \{L^1_{\ 12}=L^1_{\ 21} \},\
  ^{[1]}Q_{\beta \gamma}^\alpha =0)$$
  solve the system of $na_{(1)}$--equations (\ref{neqc1a}).
  The resulting auxiliary curved space ${V_{[1]}}$ is provided with a d--covariant differential
   operator $^{[1]}D_{\alpha }, $ defined by the d--connection
   $^{[1]}\Gamma _{\beta \tau}^\alpha =^{[1]}\gamma_{\beta \tau}^\alpha + ^{[1]}T_{\beta \tau}^\alpha.$

   The second step in the chain (\ref{chain_n1})  is parametrized by the deformations
\begin{eqnarray}
{na_{1}^{[2]}}: &{}& ^{[1]}\gamma _{\beta \tau}^\alpha =
 ^{[1]}\gamma _{\beta \tau}^\alpha + ^{[2]}P_{\beta \tau}^\alpha, \nonumber \\
 &{}& ^{[2]}T _{\beta \tau}^\alpha =
 ^{[1]}T _{\beta \tau}^\alpha + ^{[2]}Q_{\beta \tau}^\alpha, \nonumber
\end{eqnarray}
with associated values
 $$(\ ^{[2]}b_{\alpha} =({L^2_{\ 11}})^{-1}\    {}^{[1]}D_{\alpha} {L^2_{\ 11}},\
 ^{[2]}a_{\beta \gamma} =0,\ ^{[2]}P_{\beta \gamma}^\alpha = \{L^2_{\ 11} \},\
  ^{[2]}Q_{\beta \gamma}^\alpha =0 )$$
  solving the system of $na_{(1)}$--equations (\ref{neqc1a}) for fixed initial data on the auxiliary
  space ${V_{[1]}}.$
  The resulting auxiliary curved space ${V_{[2]}}$ is provided with a d--covariant differential
   operator $^{[2]}D_{\alpha }, $ defined by the d--connection
   $^{[2]}\Gamma _{\beta \tau}^\alpha =^{[2]}\gamma_{\beta \tau}^\alpha + ^{[2]}T_{\beta \tau}^\alpha.$

   The third, final, map ${na_{1}^{[3]}},$ which induces a la--spacetime ${V^{(2+2)}}$
    with the d--con\-nec\-ti\-on (\ref{cdconm})
     and d--torsions    (\ref{dtorsionsm}), could be treated as trivial $na_{(1)}$--map
     with a simple deformation of the torsion structure
  $$  ^{[3]}T _{\beta \tau}^\alpha =
 ^{[2]}T _{\beta \tau}^\alpha + ^{[3]}Q_{\beta \tau}^\alpha, $$
      which is given by the set of values
     $(\ ^{[3]}b_\alpha = 0,\  ^{[2]}a_{\beta \gamma} =0,\ ^{[2]}P_{\beta \gamma}^\alpha = 0,\
  ^{[2]}Q_{\beta \gamma}^\alpha = T_{\beta \gamma}^\alpha),$ where $T_{\beta \gamma}^\alpha$
  has just the components (\ref{dtorsionsm}).

  So, we have proved that the vacuum Einstein fields given by a metric (\ref{ansatz2}) (equivalently,
  by a d--metric (\ref{diag}), induced by a Finsler like metric, are nearly conformaly flat, being
   related ba chain of two $na_{(1)}$--maps and a deformation of torsion structure with the Minkowski
    spacetime. On every spacetime, on the initial ${V_{[0]}},$ two auxiliary, ${V_{[1]}},$ and ${V_{[2]}},$
    and on the final image, ${V^{(2+2)}}$ one holds $na_{(1)}$--invariant conditions of type
    (\ref{nainv1}).

\subsection{Chains of $na_{(2)}$--maps}
The considered vacuum la--spacetimes  could  be also induced by a
chain of three $na_{(2)}$--maps from the Minkowski spacetime,
\begin{equation}
{V_{[0]}}\ {\buildrel {na_{2}^{[1]}} \over \longrightarrow}\  {V_{[1]}}\ %
{\buildrel {na_{2}^{[2]}} \over \longrightarrow}\ {V_{[2]}}\
{\buildrel {na_{2}^{[3]}} \over \longrightarrow}\ {V^{(2+2)}}.
\label{chain_n2}
\end{equation}

The first $na_{2}$--map from  this chain  is defined by some
deformations of the symmetric part (\ref{simdef}),
 with the deformation d--tensor parametrized as (\ref{defcon2a}),
 and antisymmetric (\ref{torsdef}) part of the d--connection
\begin{eqnarray}
{na_{2}^{[1]}}: &{}& ^{[1]}\gamma _{\beta \tau}^\alpha =
 ^{[0]}\gamma _{\beta \tau}^\alpha + ^{[1]}\sigma _\beta\ ^{[1]}F_{ \tau}^\alpha, \nonumber \\
 &{}& ^{[1]}T _{\beta \tau}^\alpha =
 ^{[0]}T _{\beta \tau}^\alpha + ^{[1]}Q_{\beta \tau}^{\alpha}, \nonumber
\end{eqnarray}
where it is considered that on the flat background space is chosen
a system of coordinates for which $^{[0]}\gamma _{\beta
\tau}^\alpha =  0$ and $^{[0]}T _{\beta \tau}^\alpha = 0. $ The
values $$( {}^{[1]}{F_{\tau}^{\alpha}}=\{ L^1_{\ 12} \},
{}^{[1]}{\mu}_{\alpha} = {\partial}_\alpha \ln{|L^1_{\ 12}|},\
 {}^{[1]}{\nu}_{\beta } =0,\ {}^{[1]}{\sigma}_{\alpha} = {\delta}_{\alpha}^1,\ {}^{[1]}{Q_{\beta \gamma}^{\alpha}}=0 )$$
  solve the system of $na_{(2)}$--equations (\ref{neqc2a}).
  The resulting auxiliary curved space ${V_{[1]}}$ is provided with a d--covariant differential
   operator $^{[1]}D_{\alpha }, $ defined by the d--connection
   $^{[1]}\Gamma _{\beta \tau}^\alpha =^{[1]}\gamma_{\beta \tau}^\alpha .$

 The second $na_{2}$--map from (\ref{chain_n2}) is parametrized as
 \begin{eqnarray}
{na_{2}^{[2]}}: &{}& ^{[2]}\gamma _{\beta \tau}^\alpha =
 ^{[1]}\gamma _{\beta \tau}^\alpha + ^{[2]}\sigma _\beta\ ^{[2]}F_{ \tau}^\alpha, \nonumber \\
 &{}& ^{[2]}T _{\beta \tau}^\alpha =
 ^{[1]}T _{\beta \tau}^\alpha + ^{[2]}Q_{\beta \tau}^{\alpha}, \nonumber
\end{eqnarray}
with the values $$( {}^{[2]}{F_{\tau}^{\alpha}}=\{ L^2_{\ 11} \},\
 {}^{[2]}{\mu}_{\alpha} = ({L^2_{\ 11}})^{-1}\    {}^{[1]}D_{\alpha} {L^2_{\ 11}},\
 {}^{[2]}{\nu}_{\beta } =0,\ {}^{[2]}{\sigma}_{\alpha} = {\delta}_{\alpha}^1,\ {}^{[2]}{Q_{\beta \gamma}^{\alpha}}=0 )$$
  solving the system of $na_{(2)}$--equations (\ref{neqc2a}).
  The second resulting auxiliary curved space ${V_{[2]}}$ is provided with a d--covariant differential
   operator $^{[2]}D_{\alpha }, $ defined by the d--connection
   $^{[2]}\Gamma _{\beta \tau}^\alpha =^{[2]}\gamma_{\beta \tau}^\alpha .$

    The third step in the chain (\ref{chain_n2}) is a trivial $na_{2}$--map with ´pure´ deformation of
   d--torsions given by the values
   $$( {}^{[3]}{F_{\tau}^{\alpha}}=0,\ {}^{[3]}{\mu}_{\alpha} = 0,\  {}^{[3]}{\nu}_{\beta } =0,\
   {}^{[3]}{\sigma}_{\alpha} = 0,\ {}^{[3]}{Q_{\beta \gamma}^{\alpha}}={T_{\beta \gamma}^{\alpha}} ),$$
   where the d--torsions are those from (\ref{dtorsionsm}). We can conclude that a vacuum Einstein
    ${V^{(2+2)}}$ spacetime
   provided with a Finsler like induced metric of type (\ref{ansatz2}) (equivalently,
  by a d--metric (\ref{diag}), could be alternatively induced by a chain of $na_{2}$--maps for which,
   on every stape, one holds the invariant conditions (\ref{nainv2}). This is a particular
  property of this class of d--metrics. We shall prove in the next subsection that in a similar fashion
  we can consider chains of $na_{3}$--mapa for inducing such types of vacuum la--spacetimes.

\subsection{Chains of $na_{(3)}$--maps}
Finaly, we elucidate the possibility of inducing vacuum Finsler
like induced Einstein spaces by using chains
 of $na_{(3)}$--maps,
\begin{equation}
{V_{[0]}}\ {\buildrel {na_{3}^{[1]}} \over \longrightarrow}\  {V_{[1]}}\ %
{\buildrel {na_{3}^{[2]}} \over \longrightarrow}\ {V_{[2]}}\
{\buildrel {na_{3}^{[3]}} \over \longrightarrow}\ {V^{(2+2)}}.
\label{chain_n3}
\end{equation}

Now, the first $na_{3}$--map  is defined by some deformations of
the symmetric part (\ref{simdef}),
 with the deformation d--tensor parametrized as (\ref{defcon3a}),
 and antisymmetric (\ref{torsdef}) part of the d--connection
\begin{eqnarray}
{na_{3}^{[1]}}: &{}& ^{[1]}\gamma _{\beta \tau}^\alpha =
 ^{[0]}\gamma _{\beta \tau}^\alpha + ^{[1]}\sigma _{\beta \tau}\ ^{[1]}\varphi ^\alpha, \nonumber \\
 &{}& ^{[1]}T _{\beta \tau}^\alpha =
 ^{[0]}T _{\beta \tau}^\alpha + ^{[1]}Q_{\beta \tau}^{\alpha}, \nonumber
\end{eqnarray}
where it is considered that on the flat background space is chosen
a system of coordinates for which $^{[0]}\gamma _{\beta
\tau}^\alpha =  0$ and $^{[0]}T _{\beta \tau}^\alpha = 0. $ The
values $$( {}^{[1]}{P_{\tau \beta}^{\alpha}}=
 \{^{[1]}\sigma _{12} \ {}^{[1]}{\varphi}^{1}= L^1_{\ 12} \},\
  {}^{[1]}{\mu}_{\alpha} = {\partial}_\alpha \ {}^{[1]}{\varphi}^{1}\
 {}^{[1]}{\nu} =0,\ {}^{[1]}{\varphi}^{\alpha} = {\delta}^{\alpha}_1,\ {}^{[1]}{Q_{\beta \gamma}^{\alpha}}=0 )$$
  solve the system of $na_{(3)}$--equations (\ref{neqc3a}).
  The resulting auxiliary curved space ${V_{[1]}}$ is provided with a d--covariant differential
   operator $^{[1]}D_{\alpha }, $ defined by the d--connection
   $^{[1]}\Gamma _{\beta \tau}^\alpha =^{[1]}\gamma_{\beta \tau}^\alpha .$

The second $na_{3}$--map from (\ref{chain_n3}) is stated by
\begin{eqnarray}
{na_{3}^{[2]}}: &{}& ^{[2]}\gamma _{\beta \tau}^\alpha =
 ^{[1]}\gamma _{\beta \tau}^\alpha + ^{[2]}\sigma _{\beta \tau}\ ^{[1]}\varphi ^\alpha, \nonumber \\
 &{}& ^{[2]}T _{\beta \tau}^\alpha =
 ^{[1]}T _{\beta \tau}^\alpha + ^{[2]}Q_{\beta \tau}^{\alpha}, \nonumber
\end{eqnarray}
with the values $$( {}^{[2]}{P_{\tau \beta}^{\alpha}}=
 \{^{[2]}\sigma _{11} \ {}^{[2]}{\varphi}^{2}= L^2_{\ 11} \},\
  {}^{[2]}{\mu}_{\alpha} = ^{[2]}D_{\alpha } \ {}^{[1]}{\varphi}^{2}\
 {}^{[2]}{\nu} =0,\ {}^{[2]}{\varphi}^{\alpha} = {\delta}^{\alpha}_2,\ {}^{[2]}{Q_{\beta \gamma}^{\alpha}}=0 )$$
  defining a solution of the system of $na_{(3)}$--equations (\ref{neqc3a}).

The third step in the chain (\ref{chain_n3}) should be treated as
 a trivial $na_{3}$--map when all parameters and deformations vanishes
 exepting a ´pure´ deformation of  d--torsions,
   $ {}^{[3]}{Q_{\beta \gamma}^{\alpha}}={T_{\beta \gamma}^{\alpha}},$
   where the d--torsions are those from (\ref{dtorsionsm}).

\section{Outline of the Results}

In this paper we showed how the Einstein equations can be written
with respect to anholonomic frames with associated nonlinear
connection structures, when the dynamics of gravitational and
matter field interactions is described by mixed sets of holonomic
and anholonomic variables.

We demonstrated that by using anholonomic frames on (pseudo)
Riemannian
 spacetimes we can model  locally anisotropic interactions and  structures
  (Finsler like and more general ones)  which are defined in the
   framework of the general relativity theory. The very important  role of
   definition of frame systems in Einstein gravity was emphasized
   by explicit examples when the tetrad coefficients are selected
   as to define new classes of solutions, with generic local anisotropy,
   of the Einstein equations. There were considered vacuum and non--vacuum
   gravitational fields   induced by some generalized Finsler like metrics

   Our principial aim in this work was to underline locally anisotropic
   gravitational effects and give them a rigorous geometrical spacetime description.

   We elaborated the theory of nearly autoparallel locally anisotropic maps which generalizes
   the geometry of conformal and geodesic transforms and applied it for
    definition of conservation laws (via tensor
   integrals  and/or by introducing nearly autoparallel backgrounds)
    on spacetimes provided with anholonomic structures. Nearly autoparallel
     chain resolutions of Finsler like induced  vacuum Einstein fields
     were constructed.

 \vskip10pt {\bf Acknoweledgments} \vskip5pt S. V. work was supported by the
 German Academic Exchange Service (DAAD).


\end{document}